\documentclass[review]{elsarticle}
  \usepackage{longtable} 
\usepackage{color}
\usepackage{array}
\usepackage{tabularx}
\usepackage{tabulary}
\usepackage{multirow}
\usepackage{vmargin}
\usepackage{pifont}
\usepackage{afterpage}
\usepackage{fancyhdr}
\usepackage[mathcal]{euscript}
\usepackage{rotating}
\usepackage{amsmath,amsfonts,amsthm,bm,amssymb}
\usepackage{pdflscape}
\usepackage{booktabs}
\usepackage{siunitx}
\usepackage{latexsym}
\usepackage{graphicx}
\usepackage{enumerate}
\usepackage{scrextend}
\usepackage{algorithm,algorithmic}
\usepackage{booktabs}
\usepackage{pstricks-add}

\newtheorem{theorem}{Theorem}

\newtheorem{definition}{Definition}

\newcommand{\be}{\begin{equation}}
\newcommand{\ee}{\end{equation}}
\newcommand{\bea}{\begin{eqnarray}}
\newcommand{\eea}{\end{eqnarray}}
\newcommand{\bean}{\begin{eqnarray*}}
\newcommand{\eean}{\end{eqnarray*}}

\long\def\delete#1{}

\def\d{\delta}

\begin{document}

\begin{frontmatter}

\title{Interdependent evolution of robustness, force transmission and damage in a heterogeneous quasi-brittle granular material:\\ from suppressed to cascading failure}

\author[mymainaddress]{Antoinette Tordesillas\corref{mycorrespondingauthor}}
\cortext[mycorrespondingauthor]{Corresponding author}
\ead{atordesi@unimelb.edu.au}
\author[mymainaddress]{Sanath Kahagalage}
\author[mymainaddress]{Charl Ras}
\author[mysecondaryaddress]{Micha{\l} Nitka}
\author[mysecondaryaddress]{Jacek Tejchman}
\address[mymainaddress]{School of Mathematics $\&$ Statistics, The University of Melbourne, Australia}
\address[mysecondaryaddress]{Faculty of Civil and Environmental Engineering, Gda{\'n}sk University of Technology, Poland}

\begin{abstract}

A heterogeneous quasi-brittle granular material can withstand certain levels of internal damage before global failure.  This robustness depends not just on the bond strengths but also on the topology and redundancy of the bonded contact network, through which forces and damage propagate. Despite extensive studies on quasi-brittle failure, there still lacks a unified framework that can quantify the interdependent evolution of robustness, damage and force transmission.  Here we develop a framework to do so.  It is data-driven, multiscale and relies solely on the contact strengths and topology of the contact network for material properties.  Using data derived from discrete element simulations of concrete specimens under uniaxial tension, we uncover evidence of an optimized force transmission, characterized by two novel transmission patterns that predict and explain damage propagation from the microstructural to the macroscopic level.  The first comprises the optimized flow routes: shortest possible paths that can transmit the global transmission capacity.  These paths reliably predict tensile force chains.  The second are the force bottlenecks.  These provide an early and accurate prediction of the ultimate pattern, location and interaction of macrocracks.  A two-pronged cooperative mechanism among bottlenecks, enabled by redundancies in transmission pathways, underlies robustness in the pre-failure regime.  Bottlenecks take turns in accommodating damage, while member contacts spread the forces to confine damage to low capacity contacts which leave behind a web of strong contacts to support and curtail the failure of tensile force chains in the region.  This cooperative behavior, while serving to minimize the inevitable reduction in global transmission capacity, progressively heightens the interdependency among these contacts and elicits the opposite effect.  Ultimately, the dominant bottleneck becomes predisposed to cascading failure which, in turn, triggers abrupt and catastrophic failure of the system.\\

\end{abstract}


\end{frontmatter}

\section{Introduction}
\label{Introduction}

Studies from across material science and engineering have attributed the apparent similarities in the strength and failure of many everyday materials like sand, cereal, concrete, rocks, ceramics, ice, gels etc. to a common internal structure: an endoskeleton of interconnected grains \cite{burnley2013,jiang2014bond,bergantz2017kinematics,POLOJARVI201512,Morales2016,SUN201769}. 
Two features of this granular skeleton have received significant attention: structural and functional. The former has been characterized mainly with respect to the topology and anisotropies of the grain contact network (e.g., \cite{Kuhn2015,jiang2014bond,SUN201769,bergantz2017kinematics,TORDESILLAS2011265}); while most research into the latter have 
focussed on the contact forces and, in particular, force chains (e.g., \cite{radjai98,Majmudar_2005,philmagAT,qin2011numerical,jiang2014bond,TORDESILLAS2009706}).  
In a deforming sample, both features exhibit complex dynamics with a strongly coupled evolution.  This evolution is further influenced by microscale damage\footnote{In the systems studied here, failure at the microscale is solely due to contact breakage.}, which propagates in ways dependent on the contact strengths and robustness of the microstructural fabric~\cite{nitka2015,11025397620151001,suchorzewski2017discrete,suchorzewski2017b}.
Contact strength (capacity) is the maximum force that a contact can withstand before breaking.  Robustness is the ability of the material to maintain functionality (load-bearing capacity) in the presence of damage: some contacts may break without resulting in global failure.  This tolerance for damage is due to redundancies in the internal connectivity of the material (e.g., \cite{LUO2017264}).  Important advances in quantifying redundancy in granular structures with respect to several related aspects such as jamming, structural stability and statical indeterminacy have been reviewed in \cite{TORDESILLAS2011265}.  Redundancy implies the presence of multiple paths for force transfer, which crucially enable forces to be rerouted to alternative paths when damage occurs.  While there has been broad recognition of this fact and the importance of understanding these interdependencies, a holistic approach to the characterization and modeling of robustness, force transmission and damage, and their interdependent evolution, is apparently still lacking \cite{berthier2017damage}.  Furthermore, studies that explicitly address redundancy in force transmission pathways and resultant rerouting processes, the root cause of robustness, are notably missing. Overcoming these knowledge gaps is essential not just for prediction and control of mechanical performance, but also for rational design and fabrication of mechanically robust particulate materials by optimization of their microstructure (e.g., ~\cite{Gu2013,Suzuki_2016,bigdata,SCHENKER20081443}).

To elucidate some of the challenges, consider the transfer of forces at the grain contacts in disordered and dense granular media under load.  Damage disrupts the transmission of force by rendering certain paths inaccessible. In a redundant transmission system, however, multiple paths are available for flow.  Whenever damage degrades or breaks a contact\footnote{Damage to a contact can be defined as a reduction in the contact capacity. For a bonded contact, damage may take one of two forms: the bond is broken but the contact is maintained resulting in a degraded but non-zero capacity, or, both the bond and contact are broken resulting in zero capacity.}, flow may be redirected to alternative paths. Consequently previously latent paths may suddenly become important for force transfer, thus predisposing associated contacts to becoming overloaded and damaged.  Now consider this scenario at the level of individual force chains.  In particular, suppose there exist a force chain that is near its load-bearing capacity buttressed by a side neighbor through a single contact.  The failure of this critical contact may result in the collapse of not just the force chain, but also other contacts and force chains within striking distance, like toppling dominoes.  In turn, such a cascade of failures may propagate uncontrollably and precipitate catastrophic global failure.  Clearly this sequence of events demands a framework that can go beyond the standard statistics of contact forces and individual force chains.  Such a framework must be capable of accounting for all the available pathways for force transfer {\it across the scales} --- across a contact, between member contacts in individual force chains, and between all force chains and their supporting neighbors.  Here a framework to do so is proposed.   

Our framework capitalises on data science tools and approaches.  Although data science has transformed many fields such as medicine, finance, biology, social sciences, etc, its full promise in mechanics remains far from realized~\cite{KIRCHDOERFER2017622,bigdata}.  It is an important untapped resource for multiscale solid mechanics given the flourishing trove of microstructural data on heterogeneous solids --- from high-resolution imaging experiments (e.g.,~\cite{Majmudar_2005,tordesillas2015,SCHENKER20081443}) to discrete computational mechanics models (e.g., discrete element methods (DEM)\cite{nitka2015,11025397620151001,suchorzewski2017discrete,suchorzewski2017b}, lattice discrete particle method (LDPM)\cite{cusatis2003}).  Extracting useful insights from these multiscale and high-dimensional data sets presents many challenges that data science can help overcome.  To that end, we develop a data-driven framework that can address some of these challenges and thereby leverage microstructural data assets in solid mechanics, whether from experiments or physics-based models.

Our approach is multiscale, and explicitly takes into account the interdependent evolution of damage, force transmission and robustness in two- and three- dimensional systems with geometric and material heterogeneities. It uses network flow theory to extract quantitative insights from key force transmission and fracture patterns (force chains, crack interaction, failure cascades), based solely on data comprising the capacities and topology of the contact network.  We demonstrate its efficacy using data from two-dimensional DEM models of concrete specimens under quasistatic uniaxial tension \cite{nitka2015,van2002uniaxial}. We chose these tests because their pre-failure mechanics is governed by a contact network topology that changes solely from a progressive loss of contacts due to bond breakage: essentially no new contacts form from grain rearrangements prior to failure.  That is, each specimen responds to damage solely by ``rerouting" force transfer to alternative pre-existing pathways: no ``rewiring'' occurs that result in new pathways for force transmission.  This constrained evolution of the contact network presents an ideal starting point for demonstrating the essential elements of this framework in an explicit manner, while still retaining the key elements of force transfer and failure germane to quasi-brittle granular materials under uniaxial tension \cite{van2002uniaxial}.

The paper is arranged as follows. A brief summary of related past work is given in Section~\ref{sec:past}.  In Section~\ref{sec:data}, we discuss the data examined, before presenting our proposed framework in Section~\ref{sec:methods}.  Results are given in Section~\ref{sec:results} with key findings discussed in Section~\ref{sec:sum}.  We conclude in Section~\ref{sec:conclude}.


\section{Related past work}
\label{sec:past}

The framework we propose is drawn from network flow theory~\cite{AhujaNetworkFlows}.  The aim of a network flow analysis is to optimize the flow of an entity through a network, given the network topology and finite link capacities that cannot be exceeded.  This problem arises in many settings: traffic on a road network, fluid in a pipeline network, data on the Internet, electricity on a power grid etc.  In general, this process involves an evolution {\it on} and {\it of} the network; that is, the flow that takes place on the network can change the structure of the network, and vice versa.  For example, a link may break when a flow reaches or exceeds its capacity: this changes the network topology and, in turn, the flows both with respect to the paths they take and the fraction of flow allocated to each path.  Various disciplines contribute to this domain most notably: optimization, graph theory and computer science.  By far the most actively studied are the performance and resilience of complex flow networks ({\it viz.}, heterogeneous networks with inherent redundancies) in challenged environments where disruptions to transmission are the norm \cite{ESTRADA201289,west2010disrupted,DUENASOSORIO2009157}.  The problems of interest here belong to this class.

The link capacities, established from models and/or empirical measurements, quantify the different forms of disruption to the flow: disconnections and/or degraded transmission.  With these and the network topology as ``input'', a network flow analysis generates several ``output"  information including: the maximum flow that can be transmitted through the entire network (global capacity), the corresponding flows transmitted through the individual network links when the network transmits at the global capacity, and key transmission patterns such as the preferential flow paths and the {\it flow bottleneck} (i.e., a critical and vulnerable part of the network whose total capacity is equal to the global capacity).

Recently, we tested the potential of a network flow approach in the characterization and modeling of force transmission in various 2D and 3D granular systems ~\cite{tordesillas2013,Lin2014337,tordesillas2015,tordesillas2015network,kahagalage2017cuts}.  Grain-scale data came from physical experiments on natural and synthetic materials (i.e, sand using xray $\mu$CT, photoelastic particles using birefringence measurements) as well as discrete element simulations under many different loading conditions: triaxial compression, biaxial compression (constant volume, constant confining pressure), uniaxial compression, simple shear, pure shear and uniaxial tension.  Deformation occured in the presence of multiple failure mechanisms including: slip and rolling at contacts, force chain buckling, bond breakage and grain fracture.

In these prelude studies, the network nodes represented the grains, while the network links represented the contacts, bonded or unbonded.  By this definition, the link capacity corresponds to the contact strength: the maximum force that the contact can support before breaking.  Due to the general lack of {\it a priori} information on the contact strengths, various proxy models were used to estimate them based on the local topology of the grains engaged in the contact.  We also performed limited preliminary work on compressive force chains for bonded grains in ~\cite{tordesillas2015network}.  {\it Crucially, however, the interdependent evolution of damage, force transmission and robustness was not addressed in any of these studies. } Nevertheless, findings from \cite{tordesillas2013,Lin2014337,tordesillas2015,tordesillas2015network} provide a theoretical underpinning for this effort and we summarize them below along with key knowledge gaps.

It was hypothesized in \cite{tordesillas2013,Lin2014337,tordesillas2015,tordesillas2015network} that force transmission and energy dissipation in granular systems under load form an optimized process, such that patterns of self-organization in contacts and contact forces can be predicted from solutions to specific optimization problems.  That is, a framework for predicting the evolution of contacts and associated redistributions of forces through them can be broadly formulated as follows: {\it given a contact network topology and contact capacities, maximize/minimize one or more objective functions $\mathcal{F}$ subject to a set of constraints $\mathcal{C}$. } 
Results in \cite{tordesillas2013,Lin2014337,tordesillas2015} on unbonded grains showed a potential for flow bottlenecks to predict and explain the evolution of localized failure: the bottlenecks emerge early in the loading history and persist in sites where localized failure (i.e., shear band) eventually forms.  Findings in~\cite{tordesillas2015network} also suggest that the network flow approach bears potential in predicting compressive force chains, albeit the predictions depend strongly on the assumed distribution of the contact strengths.  Combined, these studies cast light on specific knowledge gaps that must be addressed in order to rationalize, quantitatively, the co-evolution of force transmission and failure in quasi-brittle granular media.  These are (a) accurate information on contact capacities, (b) characterization of the co-evolution of tensile and compressive force chains with damage, and (c) measures of robustness which fully account for redundancies in force pathways and the extent to which these redundancies are exploited in force reconfigurations in the face of damage.  To address these issues in a tractable manner, and without loss of generality, we employ data sets from heterogeneous concrete specimens under uniaxial tension.  Each specimen embodies the salient aspects of force and fracture propagation in the pre-failure regime without the added complication of new force pathways being created from grain rearrangements and compressive force chains.


\section{Data}
\label{sec:data}

The data sets come from a family of discrete element (DEM) models of fracture in concrete \cite{nitka2015,11025397620151001,suchorzewski2017discrete,suchorzewski2017b,nitka2018}.  Models in two- and three-dimensions were developed for 2-phase (aggregate, cement matrix), 3-phase (aggregate, cement matrix, interfacial transitional zones (ITZs)) to 4-phase material (aggregate, cement matrix, interfacial transitional, macro-voids), using the explicit 3D spherical, open-source DEM code YADE (\cite{kozicki2008new}, \cite{vsmilauer2010yade}). X-ray micro-computed tomography was used to ensure a realistic representation of the aggregate sizes and shapes and other geometrical properties of the meso-structure.  The performance of these models for describing fracture, fracture characteristics and size effect in concrete has been assessed under different loading conditions: bending \cite{11025397620151001,nitka2018}, uniaxial compression  \cite{nitka2015,suchorzewski2017discrete} and splitting tension \cite{suchorzewski2017b}.  Good agreement between numerical and experimental results on real concrete was achieved.  

\begin{figure}[H]
\centering
\begin{picture}(420,170)(0,0)
  \put(0,0){\includegraphics[width=.48\textwidth,clip]{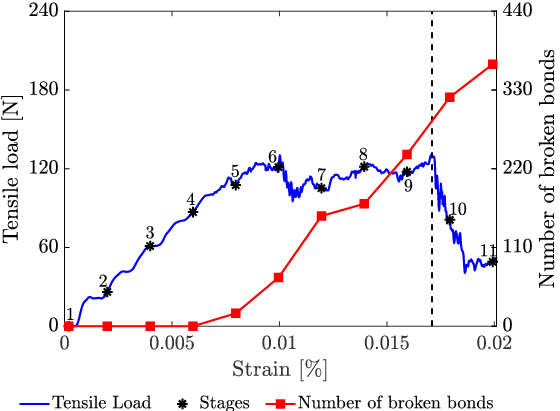}}
   \put(30,95){\includegraphics[width=.15\textwidth,clip]{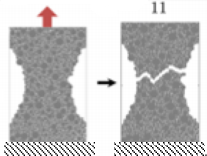}}
   \put(215,0){\includegraphics[width=.48\textwidth,clip]{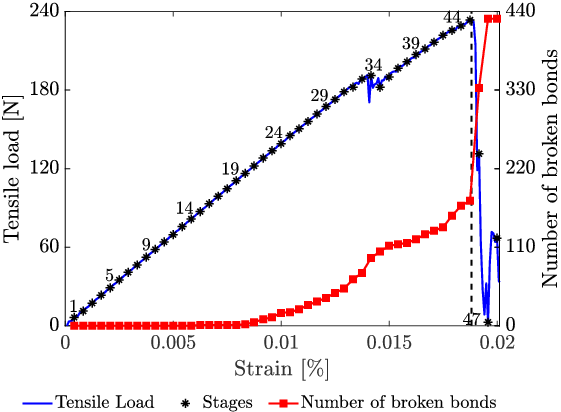}}
   \put(245,95){\includegraphics[width=.145\textwidth,clip]{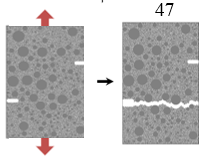}}
  \put(-5,160){\small{(a)}}
  \put(210,160){\small{(b)}}
\end{picture}
\caption{(Color online) Evolution with strain of the macroscopic tensile load and the total number of broken bonds in the specimen for (a) Data I, (b) Data II. Dashed vertical line marks the stage at peak load.}
\label{fig:bb}
\end{figure}

Here we confine our analysis to Data I and II from two concrete specimens under quasistatic, two-dimensional uniaxial tension (Figure~\ref{fig:bb}).  Laboratory tests were conducted to calibrate these models~\cite{nitka2015,11025397620151001}, following earlier experiments \cite{van2000experimental}.  Data I is from a ``dog-bone'' shaped concrete specimen: 150 mm high and 100 mm wide (60 mm at the mid-height). It consists of 4942 spherical grains: 704 aggregate grains (diameter range $2-10$ mm) and 4238 cement matrix grains (diameter range $0.5-2$ mm).  
Data II is from a rectangular concrete specimen: 150 mm high and 100 mm wide with two diagonally opposite U-shaped notches.  Each notch is of size 15 mm $\times$ 5 mm: the notch on the left (right) boundary is 50 mm (100 mm) from the bottom boundary.   The specimen is modelled as a 3-phase material composed of aggregate, cement matrix and ITZs.  The ITZs, the weakest phase, are weaker by 30\% than the cement matrix, following~\cite{xiao2013}. There are 200 aggregate spherical grains (diameter range $2-16$ mm) and 8,000 cement matrix spherical grains (diameter range $0.25-2$ mm).  Aggregate grains possess ITZs which are simulated as contacts between aggregate and cement matrix grains.  The cement matrix grains have no ITZs.  Overall, the grains comprise $95\%$ of the specimen in Data I and II.

The input data for our proposed network flow framework consist of the contact capacities and the evolving topology of the bonded contact network. For brevity, we focus the discussion below to these two aspects alone: full details are published elsewhere~\cite{nitka2015,11025397620151001}.

\subsection{Contact capacities}  
Albeit bonds can break by shear in both specimens, the essential microscale mechanism for damage in the pre-failure regime is bond breakage in tension~\cite{nitka2015}.  No new bonded contacts form during the tests.  Moreover, bond breakage immediately results in loss of contact until after peak load when a small fraction of slip contacts emerge: from a mere 0.06\% of all contacts just after peak through to 0.18\% at the residual state for both specimens.  Shear forces are relatively small throughout the pre-peak regime: at peak load, the sum of the magnitudes of the tangential contact forces relative to that of the total contact forces is 26\% for Data I (17\% for Data II).  

Thus the contact capacity that is relevant to our analysis is the tensile bond strength, which is a function of the minimum tensile normal stress $T_n$ and the size of the grains in bonded contact.  Specifically, if the minimum normal force $F^n_{min}$ was reached, the bonded contact was broken.  Thus the contact capacity function $u$ is given by
\begin{equation}
u=F^n_{min}=T_n(r^2_{min}).
\label{eqn:cap}
\end{equation}
where $T_n$ is the minimum tensile normal stress and $r_{min}$ is the radius of the smaller of the two spherical grains in bonded contact.  
For Data I, $T_n=25$ MPa.  For Data II, $T_n=24.5$ MPa for cement-cement contacts, while $T_n=17.5$ MPa for cement-aggregate(ITZ) contacts.  Additional information on the contact laws and material parameters is given in the supplementary file.

\subsection{Evolving topology of the bonded contact network}

\begin{figure}[H]
\centering
\begin{picture}(420,330)(0,0)
  \put(0,0){\includegraphics[width=1\textwidth,clip]{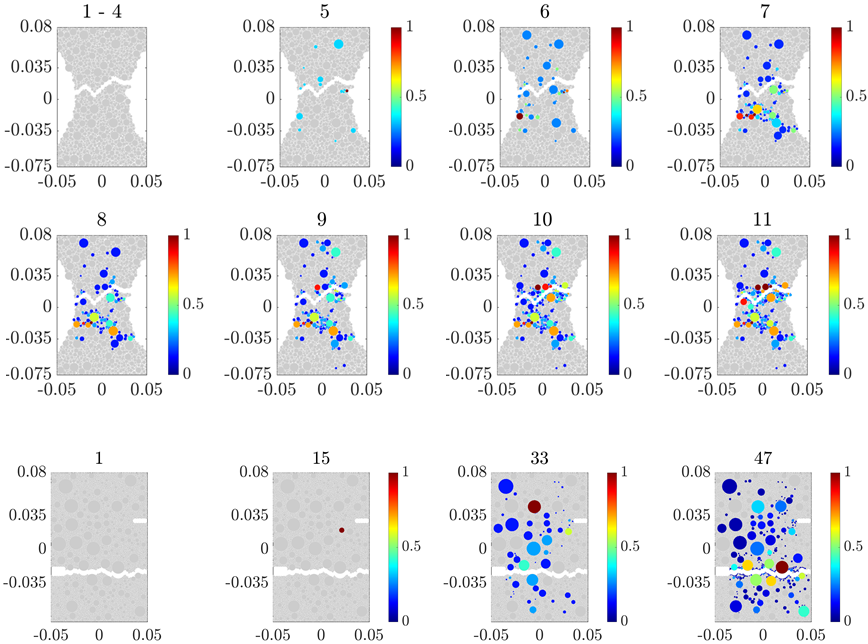}}
  \put(-5,320){\small{(a)}}
  \put(-5,95){\small{(b)}}
\end{picture}
\caption{(Color online) Spatial distribution of grain damage $d_p$ (ratio of the number of broken bonded contacts to the initial number of bonded contacts). (a) Data I and (b) Data II.  Undamaged grains, $d_p=0$, are colored gray.  Damaged grains, ${d_p}>0$, are colored according to ${\hat{d}_p}$ (${d_p}$ normalized to its maximum value for the given stage) using a blue-red colormap to enhance the contrast.  An artificial separation between the grains on either side of the macrocrack that develops after peak load is introduced to aid a visual comparison with the particles that sustained damage.  Damage below (above) the separation dominates over stages 7-8 for Data I (stages 32-33 for Data II).}
\label{fig:ngd}
\end{figure}

In the test that generated Data I, the specimen was subjected to a top boundary-driven tension while the bottom boundary was held at a fixed vertical position; both boundaries were free to move in the horizontal direction.  The constant vertical strain rate of $10^{-3}$ per second, applied on the top boundary, was small enough to ensure the test was conducted under quasistatic conditions.  The evolution of the macroscopic tensile load (vertical force on the top boundary) and the number of broken bonds (normal grain contacts) with the vertical normal strain is shown in Figure~\ref{fig:bb} (a).  We analyze 11 equilibrium stages of the loading program: stages 1-9 in the pre-peak load regime and stages 10-11 in the post-peak failure regime.   The initial and final number of bonds were 12,350 and 11,984, respectively.  Initially, bonds did not break (stages 1-4, Figure~\ref{fig:bb} (a)).  From stage 5 onwards, damage was progressive as seen in the steady increase in the population of broken bonds (reaches 240 in stage 9 just before the peak load).   Damage evolves as shown in Figure~\ref{fig:ngd} (a).  The initial sites of damage were spread throughout the specimen, although fracture started almost simultaneously at the right and left corners of the specimen along the mid-region or ``neck'' of the specimen (stage 5); these form the ends of what will later become the primary macrocrack.  At stage 6, a group of interconnected bonds on the lower left region of the specimen collectively broke, giving way to a second macrocrack in the lower region of the specimen (stages 7-8).  Just before peak load, the crack in the mid-region became dominant as damage spread rapidly across the neck of the specimen, while the second crack did not further develop (stages 9-11).  The primary macrocrack formed in stage 11, exhibiting a zig-zag pattern that spanned the neck of the specimen.

In the test that generated Data II, the top and bottom boundaries of the specimen were pulled in opposite directions, at a constant vertical strain rate of $10^{-3}$ per second; both boundaries were free to move in the horizontal direction.  We analyzed 47 equilibrium stages: stages 1-45 in the pre-peak load regime and stages 46-47 in the post-peak failure regime (Figure~\ref{fig:bb} (b)). The initial and final number of bonds were 23,492 and 23,062, respectively.  Initially, bonds did not break (stages 1-14).  From stage 15 onwards, progressive damage can be seen in the near-linear increase in the population of broken bonds (reaches 175 in stage 45 just before peak load).  Damage evolves as shown in Figure~\ref{fig:ngd} (b); see also supplementary Figure 1. The sites of damage were initially spread throughout the specimen, but began to concentrate in two regions in stages 24-33.  The first site was near the lower left notch (stage 24): just before and around peak load, damage here spread rapidly across to the right of the specimen, leading to the primary macrocrack at stage 47.   The second site, located in the weak ITZ zones in the upper region of the specimen just above and near the second notch ($\approx 0.035$ above the specimen midline), gave way to a second macrocrack at stages 32-33.  

In the pre-failure regime for both Data I and II, the following were observed for each specimen: essentially no new contacts formed, mainly tensile force chains emerged (see Figures 1-4 in \cite{DiB2018}), and patterns of fracture propagation, including multiple crack interaction and microscale failure cascades, were consistent with those in the experiments~\cite{nitka2015,11025397620151001}.

\section{Method}
\label{sec:methods}

\begin{figure}[H]
\begin{picture}(420,410)(0,0)
  \put(0,3){\centering\includegraphics[width=1\textwidth]{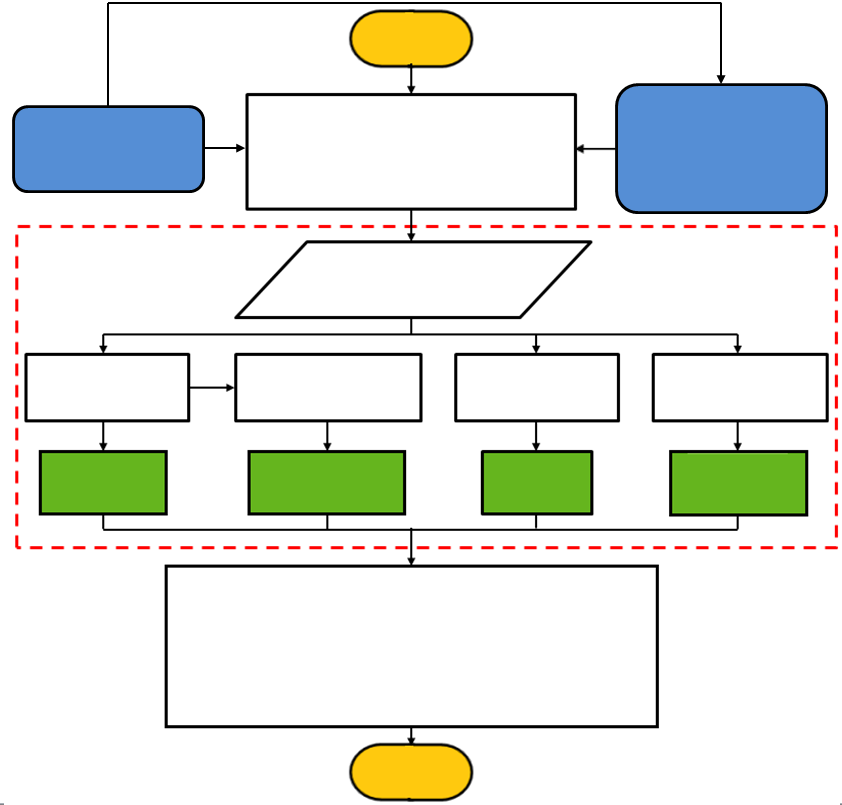}}
     \put(197,390){\textbf{Start}}
      \put(170,350){\textbf{INPUT DATA}}
      \put(128,340){$\bullet$ Bonded contact network $\mathcal{N}$}
      \put(128,325){$\bullet$ Contact capacities}
      \put(128,310){$\bullet$ Direction of applied tensile loading}
      \put(155,272){Construct Flow Network $\mathcal{F}$}
      \put(190,260){(Sec 4.1)}
      \put(18,220){Solve Maximum}
      \put(22,210){Flow Problem}
      \put(32,200){(Sec 4.1)}
      \put(133,220){Solve Minimum}
      \put(145,210){Cost Flow}
      \put(129,200){Problem (Sec 4.2)}
      \put(240,220){Solve Minimum}
      \put(244,210){Cut Problem}
      \put(250,200){(Sec 4.3)}
      \put(342,220){Solve Minimum}
      \put(355,210){Edge Cut}
      \put(338,200){Problem (Sec 4.4)}
     \put(28,170){Maximum}
     \put(35,155){Flow \textbf{$F^*$}}
     \put(130,170){Optimized Flow}
     \put(148,155){Routes $\mathcal{P}$}
     \put(252,170){Minimum}
      \put(260,155){Cut \textbf{$B$}}
      \put(355,170){Minimum}
      \put(343,155){Edge Cut \textbf{$B_\mathrm{min}$}}
     \put(200,15){\textbf{End}}
     \put(190,110){\textbf{OUTPUT}}
     \put(87,95){$\bullet$ Measure of global transmission capacity from $F^*$}
     \put(87,80){$\bullet$ Prediction of tensile force chains from $\mathcal{P}$}
      \put(87,65){$\bullet$ Prediction of crack interaction \& macrocrack from $B$}
     \put(87,50){$\bullet$ Measures of robustness from $\mathcal{P}$, $B$ \& $B_\mathrm{min}$}
     \put(20,340){\textbf{\large{Physical}}}
     \put(20,325){\textbf{\large{Experiments}}}
          \put(325,355){\textbf{\large{Discrete}}}
     \put(325,340){\textbf{\large{Computational}}}
          \put(325,325){\textbf{\large{Mechanics}}}
     \put(325,310){\textbf{\large{Models}}}
\end{picture}
\caption{Leveraging microstructural data assets on heterogeneous solids: flow chart summarizing the data-driven network flow framework for uncovering new insights on force and fracture propagation from patterns in data. \label{fig:flowchart}}
\end{figure}
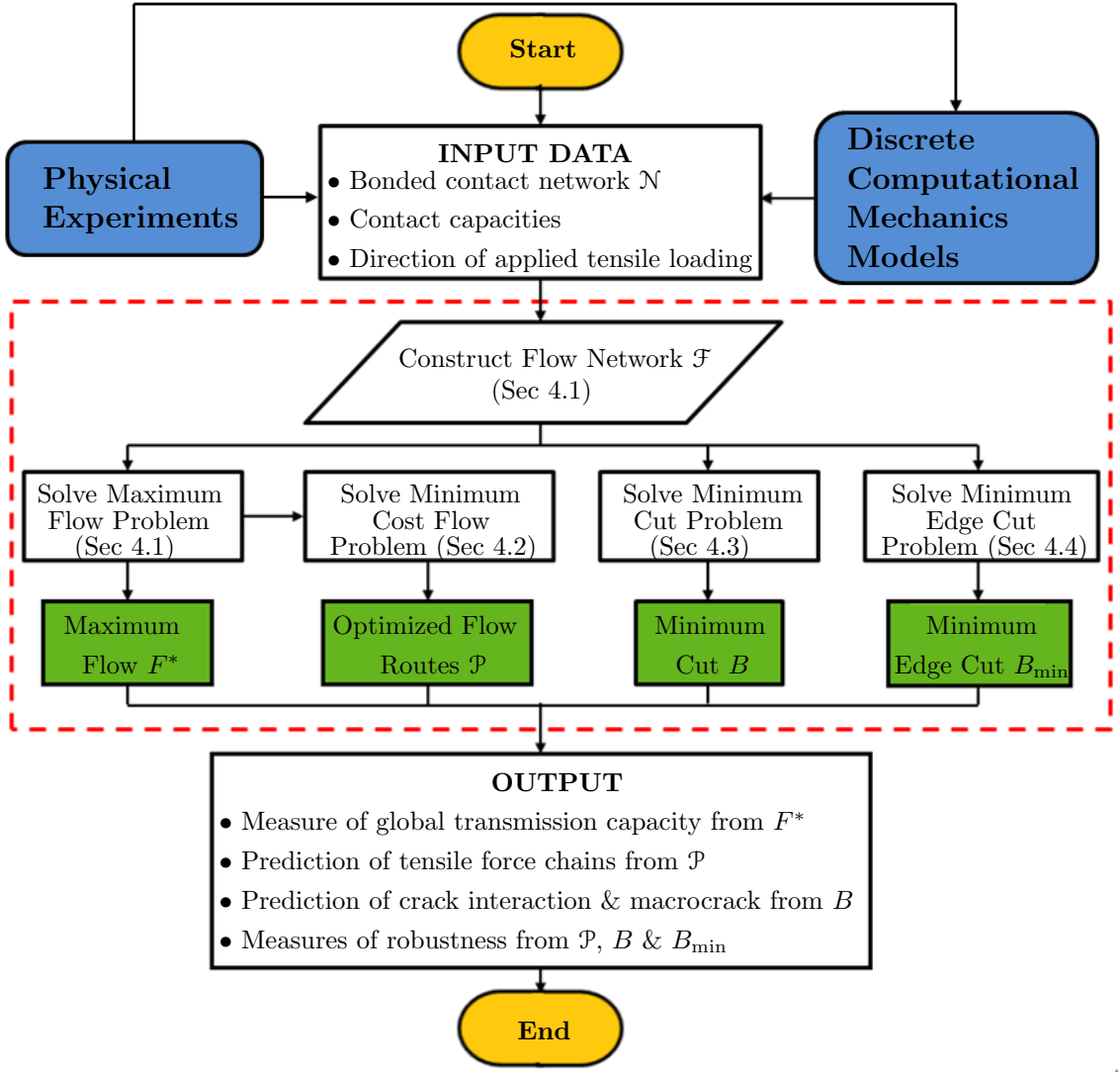

Our data-driven framework is summarized in Figure \ref{fig:flowchart}.   Using network flow theory~\cite{AhujaNetworkFlows}, this framework is designed to characterize and/or predict, from patterns in data, the following force and fracture propagation properties: (a) the transmission of normal tensile force, which we model by flow networks and maximum flow (Section~\ref{secFlow}); (b) the tensile force chains which we model by the optimized flow routes that comprise the shortest flow-pathways (Section~\ref{secPath}); (c) crack interaction and propagation to the macrocrack from the force bottlenecks which we model by the minimum cut (Section~\ref{secBot}); and (d) the transmission robustness of the contact network, which we measure by pathway redundancy and reroute score (Section~\ref{secRob}).\footnote{A list of symbols and their definitions can be found in Appendix A.}  The key idea is to represent the transmission of tensile forces in the specimen as ``flows'' through the contact network --- for the purposes of extracting nontrivial patterns in the data.  This representation is valid and fit for purpose since the physical laws that govern the transmission of vector forces in the specimen satisfy the constraints of the underlying optimization problem for the scalar flows in the flow network. 

In Section \ref{secFlow} we describe the construction of a flow network based on the bonded contact network.  Each node of this network represents a grain that can transmit tensile force to another grain, while each link represents a bonded contact.  Each link has a scalar weight which is given by the tensile strength or capacity of the corresponding bond.  Following the given data: bond breakage results in link breakage and no new links can form in the flow network as loading proceeds.  Here we describe an optimization problem which maximizes the total force flow that the flow network can sustain under the given topology and link capacities of the bonded contact network. This problem is called the \textit{Maximum Flow Problem} (MFP) and we denote the maximum flow by $F^*$.

In Section \ref{secPath} we extend the optimization problem of MFP to consider the flow pathways that are generated during maximum flow. In particular, we identify the optimized flow routes which comprise the shortest possible pathways through the network. To model this behaviour we employ the \textit{Minimum Cost Flow Problem} (MCFP).  This is an optimization problem on flow networks which maximizes flow while using the shortest available flow pathways. This set of optimized flow routes is denoted by $\mathcal{P}$.

In Section \ref{secBot} we describe a relationship between the maximum flow in a flow network, and the total capacity of all links constituting the bottleneck, also called the \textit{Minimum cut}, of the network. This relationship follows from the well-known \textit{Max-flow min-cut theorem} in network flow theory. The minimum cut is denoted by $B$.

In Section \ref{secRob} we quantify transmission robustness.  First we show that pathway redundancy, which gives the number of available transmission pathways through the specimen, can be computed from the {\it Minimum edge cut}.  This cut contains the minimum number of links, and we denote this cut by $B_\mathrm{min}$.  Second, we formulate a reroute score, a measure of the extent to which the system uses the available pathway redundancy to redirect flows to alternative paths as links are progressively broken from damage.

Finally, as shown in Figure \ref{fig:flowchart}, for a given applied tensile load, the proposed data-driven framework relies on input data which consist of the topology and bond capacities of the contact network, through which forces and damage propagate.  The output of the analysis comprises a measure of the transmission strength (derived from $F^*$), a prediction of the tensile force chains (derived from $\mathcal{P}$), an early prediction of localized failure zones and their interaction in the pre-failure regime (derived from $B$), and two measures of transmission robustness (derived from $\mathcal{P}$, $B$ and $B_\mathrm{min}$).

\subsection{Flow network $\mathcal{F}$ and maximum flow $F^*$}
\label{secFlow}

Consider the bonded contact network $\mathcal{N}$ of the granular specimen: we map each grain with at least one bonded contact to a node and each bonded contact that can transmit tensile force to a link.\footnote{Rattler grains with no contacts or grains with only unbonded contacts are excluded from $\mathcal{N}$.} Thus, both the number of nodes and the number of links in $\mathcal{N}$ vary as loading proceeds. Next we model $\mathcal{N}$ as a directed network $G$ where all links are directed.  Note that a directed link is usually called an arc. In $G$, a directed arc exists from a node $w$ (representing a grain $w$) to a node $v$ if grain $w$ is in bonded contact with grain $v$. This is clearly a symmetric relationship: arc $(w,v)$ is in $G$ if and only if arc $(v,w)$ is in $G$. The set of nodes (grains) of $G$ is denoted by $V$, and the set of arcs (bonded contacts) is denoted by $E$. Every arc $e$ is associated with a non-negative {capacity} $u_e$ which corresponds to the tensile bond strength, as governed by the capacity function in Equation~\ref{eqn:cap}. The capacity function for all of $E$ is therefore written as $$u: E \rightarrow \mathbb{R}_{+}.$$
We designate one of the nodes of $G$ as the \emph{source} $s$, and another as the \emph{sink} $t$. The quadruple $\mathcal{F} = (G, u, s, t)$ is called a {\it flow network}.

We propose a measure of global transmission capacity in the form of $F^*$, the maximum flow that the flow network $\mathcal{F}$ can sustain given its topology and link capacities.  This involves solving an optimization problem known as the \textit{Maximum Flow Problem} (MFP).  The MFP is equivalent to assigning force flows $x_e$ to every arc $e$ in $G$, without violating the two constraints of conservation of flow and the capacity rule, such that the total amount of flow transmitted out of the source (or into the sink) in $G$ is as high as possible.  Constraint 1 of MFP is the conservation of flow: this requires that for any node in $G$ except the source $s$ and the sink $t$, the sum of the flows entering the node must equal the sum of the flows leaving the node.  Constraint 2 of MFP is the capacity rule: this dictates that the amount of flow transmitted through any arc $e$ in $G$ is limited by the corresponding link capacity $u_e$.

\begin{figure}[ht]
\begin{picture}(410,200)(0,0)
  \put(5,0){\centering\includegraphics[height=.4\textwidth]{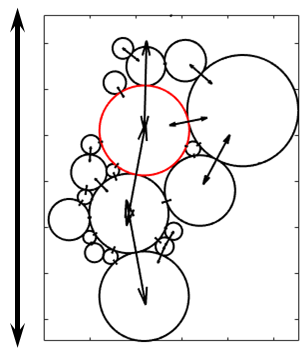}}
    \put(195,-10){\centering\includegraphics[height=.42\textwidth]{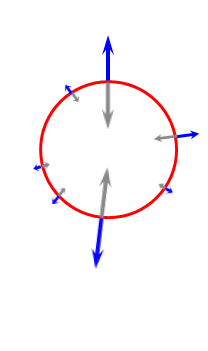}}
    \put(340,-5){\centering\includegraphics[height=.4\textwidth]{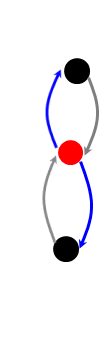}}
     \put(0,170){\small{(a)}}
     \put(185,170){\small{(b)}}
     \put(320,170){\small{(c)}}
     \put(0,10){\rotatebox{90}{Direction of applied tensile load}}
      \put(75,132){{\color{red}\Large{\textbf{1}}}}
      \put(55,65){{\color{red}\Large{\textbf{2}}}}
      \put(60,100){{\color{red}\huge{\textbf{i}}}}
      \put(230,95){{\color{red}\huge{\textbf{i}}}}
      \put(248,25){\Large{\textbf{$\text{F}_{\text{2i}}$}}}
      \put(255,145){\Large{\textbf{$\text{F}_{\text{1i}}$}}}

      \put(372,115){{\color{red}\Large{\textbf{1}}}}
      \put(372,51){{\color{red}\Large{\textbf{2}}}}
      \put(355,88){{\color{red}\huge{\textbf{i}}}}
      \put(340,160){\text{{Flow entering node i}}}
      \put(340,145){$=${\textbf{$|\text{F}_{\text{i1}}|$}}$+$ {\textbf{$|\text{F}_{\text{i2}}|$}}}
      \put(340,20){\text{{Flow leaving node i}}}
      \put(340,5){$=${\textbf{$|\text{F}_{\text{1i}}|$}}$+$ {\textbf{$|\text{F}_{\text{2i}}|$}}}
      \put(390,60){{\textbf{$|\text{F}_{\text{2i}}|$}}}
      \put(340,60){{\textbf{$|\text{F}_{\text{i2}}|$}}}
      \put(390,113){{\textbf{$|\text{F}_{\text{i1}}|$}}}
      \put(340,113){{\textbf{$|\text{F}_{\text{1i}}|$}}}
\end{picture}

\caption{(Color online) (a) A representative grain cluster embodying a tensile force chain with member grains $\lbrace1,\textbf{i},2\rbrace$, taken from Data I.  Dominant normal tensile forces on grain \textbf{i} are aligned in the direction of the applied tensile load. (b) Blue (gray) arrows correspond to forces acting on $\textbf{i}$ from neighbor grains (forces from $\textbf{i}$ acting on neighbor grains).  (c) Depiction of the conservation rule: force flow out of node \textbf{i} = force flow into node \textbf{i} (for clarity, only the arcs transmitting the dominant forces are shown).}
\label{fig:consRule}
\end{figure}
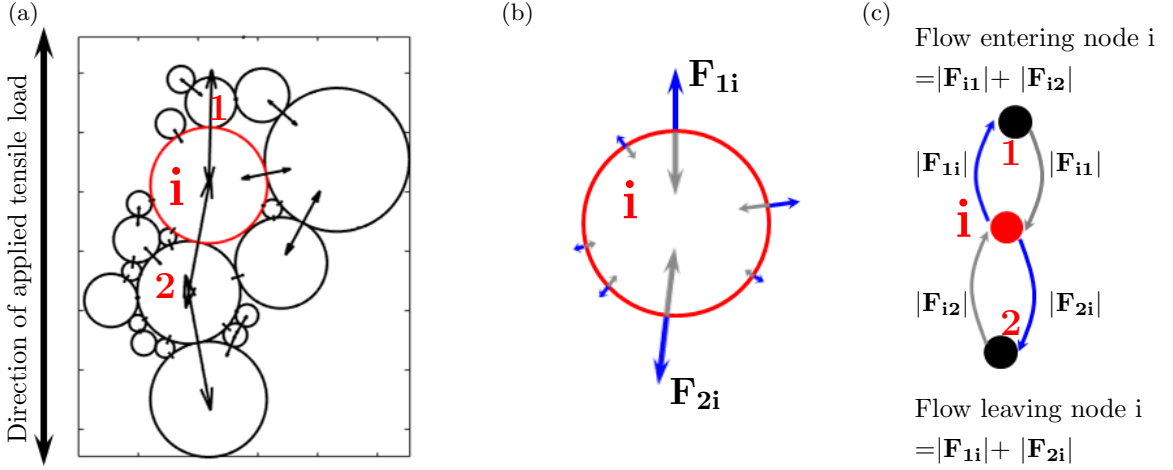

The vector forces that develop in the sample satisfy the two constraints of MFP.  The conservation rule (Constraint 1 of MFP) is satisfied due to Newton's third law.   We illustrate this in Figure~\ref{fig:consRule}(a) for a representative tensile force chain grain from Data I.  As shown in Figure~\ref{fig:consRule}(b), the force flow entering grain \textbf{i} is the sum of the magnitudes of the forces from grain \textbf{i} on its neighbors (gray vectors), while the force flow leaving grain \textbf{i} is the sum of the magnitudes of the forces acting on grain \textbf{i} from its neighbors (blue vectors).  Since grain \textbf{i} is in equilibrium, the conservation of flow rule holds: the force flow entering grain \textbf{i} is equal to the force flow leaving grain \textbf{i}.  In Figure \ref{fig:consRule}(c) we show the nodes and arcs between grain \textbf{i} and grains \textbf{1} and \textbf{2}. Note that there are two oppositely directed arcs for each contact to take account of the symmetry of grain-grain links.  The capacity rule (Constraint 2 of MFP) is satisfied by the bond breakage criterion: the tensile force in any bonded contact cannot exceed the strength or capacity of the bond (Equation~\ref{eqn:cap}).

\begin{addmargin}[1.5em]{1.5em}

\begin{definition}
\label{def:flow} {\em Given a flow network $\mathcal{F} = (G,u,s,t)$, a flow $x$ 
$$x: E \rightarrow \mathbb{R}_{+},\;e \mapsto x_e$$ is called a {\it feasible} {\em $(s,t)$-flow}, if it satisfies:

\noindent
(a) the conservation of flow
\begin{equation}
\sum_{e \in \d^{-}(v)} x_e \;= \;\sum_{e \in \d^{+}(v)} x_e, \; \; \; \forall v \in V - \{s, t\}, 
\label{eq:flow-con} 
\end{equation}
where ${e \in \d^{-}(v)}$ denotes arcs entering node $v$ and ${e \in \d^{+}(v)}$ denotes arcs leaving node $v$;
\noindent
(b) the capacity rule
\begin{equation}
0 \le x_{e} \le u_{e},\; \; \; \; \forall e \in E.
\label{eq:flow-caps}
\end{equation}

\noindent The {\em value of a flow} $x$, or net flow transmitted from the source $s$, is defined as
\begin{equation}
\label{eq:flow-vals} f(x) = \sum_{e \in \d^{+}(s)} x_e -
\sum_{e \in \d^{-}(s)} x_e.
\end{equation}
}
\end{definition}

\noindent 
We now have the following mathematical definition for our optimization problem.
\bigskip

\noindent\textbf{The Maximum Flow Problem (MFP):} Given a flow network $\mathcal{F} = (G, u, s, t)$, find a feasible $(s, t)$-flow $x$ such that $f(x)$ is maximum.

\end{addmargin}
\bigskip

In grain contact networks, it is not immediately clear how to designate the source and sink nodes ($s$ and $t$). Since the applied tensile load is in the vertical direction, an intuitive approach is to select all nodes in contact with the top wall as sources, and all nodes in contact with the bottom wall as sinks. However, this results in a multiple source and sink flow problem, which our definition (above) does not incorporate. Fortunately, the multiple source and sink problem can be reduced to the classical maximum flow problem. We do this by selecting the top and bottom walls as two new nodes of $G$, i.e., the supersource and supersink, respectively; and then assigning infinite capacity to all contacts between walls and grains. In mathematical formalism we have the following:\\

\begin{addmargin}[1.5em]{1.5em}
Let $S\subset V$ and $T\subset V$ be disjoint sets of nodes of $G$. The quadruple $(G, u, S,
T)$ is called a \emph{multiple-source, multiple-sink flow network}.

\bigskip
\begin{definition}
\label{def:flowM} {\em Given a multiple-source multiple-sink flow network $(G, u, S, T)$, a flow
$$x: E \rightarrow \mathbb{R}_{+},\;e \mapsto x_e$$ is called a {\em feasible $(S,T)$-flow} if it satisfies:
\begin{eqnarray}
\sum_{e \in \d^{-}(v)} x_e = \sum_{e \in \d^{+}(v)} x_e,\; \; \; \forall v \in V - (S\cup T); \label{eq:flow-cons} \\
0 \le x_{e} \le u_{e},\;\; \;  \forall e \in E. \label{eq:flow-cap}
\end{eqnarray}
The {\em value of a flow} $x$, or net flow transmitted from the supersource $s$, is defined as
\begin{equation}
\label{eq:flow-val} f(x) = \sum_{e \in \d^{+}(S)} x_e -
\sum_{e \in \d^{-}(S)} x_e.
\end{equation}
}
\end{definition}

\noindent To solve the multiple-source multiple-sink maximum flow problem, we convert it to a (single source, single sink) Maximum Flow Problem as follows. Given $(G, u, S,T)$ create a new node, the supersource $s$, and add an arc from $s$ to every node in $S$. Similarly, create a new node, the supersink $t$,  and add an arc from every node in $T$ to $t$. Let the capacity of all new arcs be infinite. It can easily be shown that a maximum flow on the resulting flow network is a maximum flow on $(G, u, S,T)$ when restricted to the arcs of $G$.  Even for large networks, this problem can be solved efficiently by the use of existing algorithms~\cite{AhujaNetworkFlows}.

\end{addmargin}

\bigskip

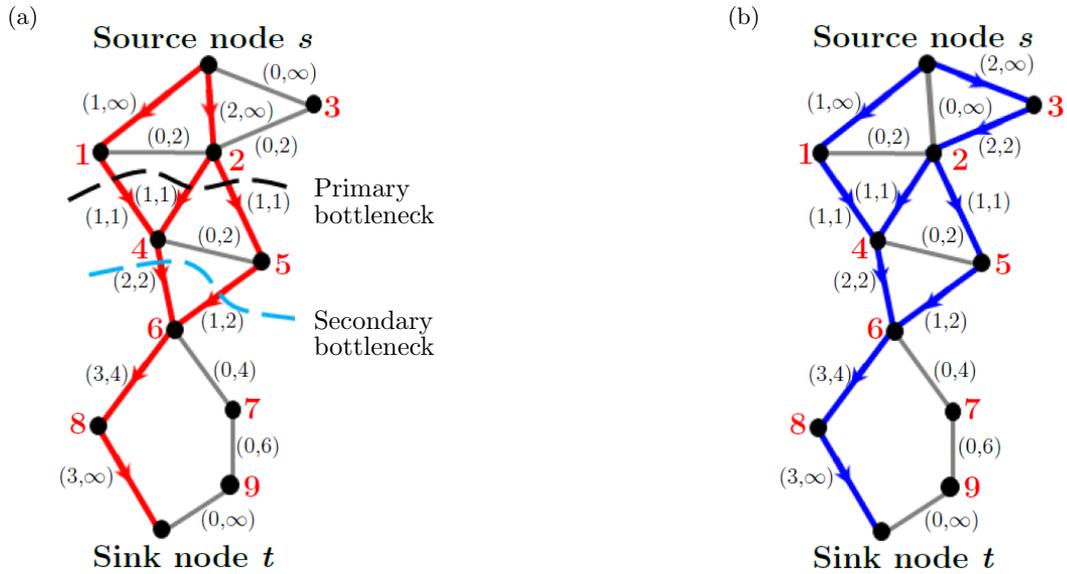
\begin{figure}[ht]
\begin{picture}(420,210)(0,0)
  \put(20,0){\centering\includegraphics[height=0.5\textwidth]{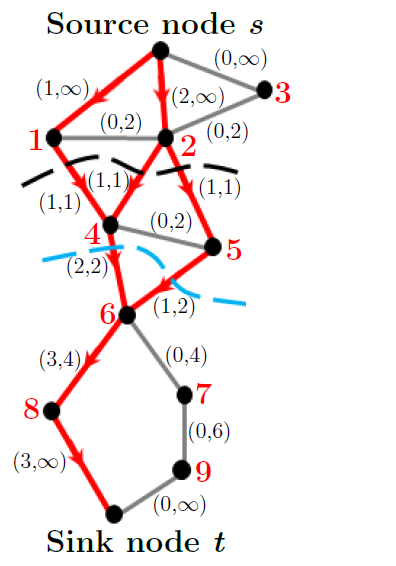}}
   \put(280,0){\centering\includegraphics[height=0.5\textwidth]{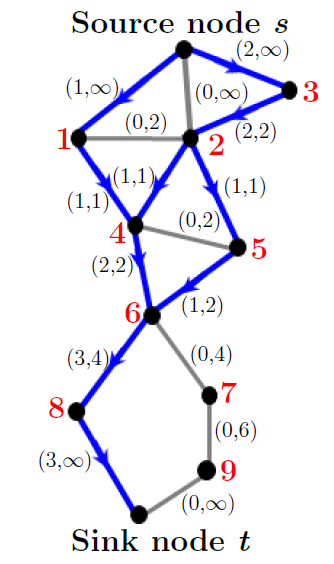}}
   \put(120,145){Primary}
   \put(120,135){bottleneck}
   \put(120,95){Secondary}
   \put(120,85){bottleneck}
  \put(5,210){\small{(a)}}
  \put(275,210){\small{(b)}}
\end{picture}
\caption{(Color online) A depiction of the multiple concepts discussed in Sections 4.1-4.4.  Two examples of a feasible $(s,t)$-flow with maximum flow $F^*$ = 3. Link labels are of the form $(x_e,u_e)$. (a) Maximum flow pathways with unit cost $\mathcal{P}$ (red arcs), primary bottleneck or minimum cut (black dashed line), and secondary bottleneck (light blue dashed line). (b) Maximum flow pathways without cost (blue arcs). Gray arcs in (a) and (b) are not part of the maximum flow pathways: arcs with zero flow.} 
\label{fig:mf}
\end{figure}

Figure~\ref{fig:mf} is a toy example of a flow network showing two different flow assignments of equal value.  Note that this figure serves to illustrate multiple aspects of the framework, both here and in future sections.  Of relevance to the present discussion are the added arcs with infinite capacity from supersource $s$ to nodes in $S$ and from nodes in $T$ to supersink $t$, where set $S$ is $\lbrace1, 2, 3\rbrace$ and set $T$ is $\lbrace8, 9\rbrace$. We use labels $(x_{e},u_{e})$ for each link $e$ such that: $x_e$ is the flow on $e$, $u_e$ is the capacity of $e$, and each undirected link in the figure represents two arcs in opposite directions. The directed arcs which are not shown here correspond to zero flow. The maximum flow value $F^*$ for this example is $3$, as can be seen by summing the flows entering $t$ or leaving $s$.

\bigskip

\subsection{The optimized flow routes $\mathcal{P}$ }
\label{secPath}

As seen from Figure~\ref{fig:mf}, solutions to the MFP are generally not unique. Depending on the algorithm used, different pathways may be used to transmit the flow through $\mathcal{N}$ from source to sink. In grain contact networks, there is a tendency for the preferred force transmission pathways --- the tensile force chains --- to self-organize in quasilinear formation in alignment with the direction of the major principal stress axis (the vertical direction for the specimens studied here).  With respect to $\mathcal{N}$, such a pattern of self-organization implies relatively short transmission paths. To capture this behavior we extend our model to the \textit{Minimum Cost Flow Problem} (MCFP). Solutions to this problem maximize total flow while also using paths that are as short as possible. This is achieved by associating a unit cost with every arc in the network, and then solving the optimization problem of finding a maximum flow through the network while also minimizing the total cost of all arcs used in the maximum flow. As with the MFP, there exist algorithms that can solve the MCFP for large networks~\cite{AhujaNetworkFlows}. 

In Figure~\ref{fig:mf}, we depict two different maximum flow pathways: (a) maximum flow pathways with unit cost $\mathcal{P}$ (red arcs); (b) maximum flow pathways without cost (blue arcs). In Figure~\ref{fig:mf}, gray arcs are not utilized in the flow pathways. As can be seen from Figure~\ref{fig:mf} (a), the optimized routes $\mathcal{P}$ comprises the flow pathways $s-1-4-6-8-t$, $s-2-4-6-8-t$, and $s-2-5-6-8-t$.  Each pathway has a length of $5$ (5 arcs). The maximum flow value is $3$ and the total cost (with unit cost for each arc) is $15$. In Figure~\ref{fig:mf} (b) the flow pathways are $s-1-4-6-8-t$, $s-3-2-4-6-8-t$, and $s-3-2-5-6-8-t$. These paths have lengths 5, 6, and 6 and the maximum flow value is also $3$. It is clear that $\mathcal{P}$ utilizes shorter paths than the flow paths without cost.  Note also that $\mathcal{P}$ \emph{is not necessarily a subset} of the paths used in the MFP. From Figure~\ref{fig:mf}, it is clear that link $(s,2)$ is in $\mathcal{P}$, but not used in the flow of Figure~\ref{fig:mf} (b). We now provide a formal definition of the MCFP.

\bigskip

\begin{addmargin}[1.5em]{1.5em}
\noindent\textbf{The Minimum Cost Flow Problem (MCFP):}
The input of MCFP is a quadruple $(G, u, b, c)$, where $G = (V, E)$ is a directed graph,

$$u: E
\rightarrow  \mathbb{R}_{+} \cup \{\infty\}$$
is a \textit{capacity} function on arcs of $G$,
$$b: V \rightarrow  \mathbb{R}$$
is a \emph{demand} function satisfying
$$
\sum_{v \in V} b_v = 0,
$$
and
$$c: E \rightarrow  \mathbb{R}$$
is a {\em cost} function.
We call $b_v$ the {\em demand} (or {\em balance}) of $v$ and $c_e$ the {\em cost} of $e$. If $b_v \ge 0$ then $v$ is a ``demander"; if $b_v < 0$ then $v$
is a ``supplier".  Based on these, and the capacity $u_e$ of arc $e$, the minimum cost flow problem is defined as follows. 

\bigskip
\noindent
We minimize the function $$\sum_{e \in E} c_e x_e,$$ subject to the net flow transmitted from any node $v$ being given by
\begin{equation}
\sum_{e \in \d^{-}(v)} x_e - \sum_{e \in \d^{+}(v)} x_e = b_v,\; \;\; \;  \forall v \in V, 
\label{eq:min-cost-bl}
\end{equation}
and
\begin{equation}
0 \le x_{e} \le u_{e},\; \;\;   \forall e \in E. 
\label{eq:min-cost-fsb}
\end{equation}

\noindent 
In our model, we let $c_e=1$ for all arcs $e$; we let $b_v=0$ for all $v\neq s$ and $v\neq t$; and we let $b_s=-F^*$ and $b_t=F^*$ where $F^*$ is the maximum flow in $G$. Since the cost $c_e$ of every arc is the same, solving the MCFP with these parameters creates an $(s,t)$-flow of value $F^*$ while using flow paths that have as few arcs as possible.

\end{addmargin}

\subsection{The minimum cut and the flow bottleneck $B$}
\label{secBot}

Next we look at so-called ``cuts" of a network. A \textit{cut} of a flow network is simply a partition of the nodes of the network so that the source is in one part of the partition and the sink in another. The capacity of a cut is the sum of the capacities of all arcs that leave the set of nodes containing the source and enter the set of nodes containing the sink. The cut with the minimum capacity is called the {\it minimum cut}.  The minimum cut identifies the flow {\it bottleneck}.  The bottleneck is highly prone to congestion and is thus the most vulnerable part of a transmission network.

In a grain contact network, a cut can be visualized as a literal partition of the set of grains by means of a crack. In this sense the capacity of the cut is the sum of the contact strengths (here bond strengths) between the grains separated in the crack.  In Figure \ref{fig:mf}(a) we depict two distinct cuts in a contact network using dashed lines. The first cut is the minimum cut (the primary bottleneck), highlighted by the black dashed curve.  It partitions the set of grains into two sets $\lbrace1, 2, 3\rbrace$ and $\lbrace4, 5, 6, 7, 8, 9\rbrace$.  The links in this cut are $(1,4)$, $(2,4)$, and $(2,5)$, each with a capacity of 1. The total capacity of this cut is therefore $3$. The second cut, highlighted by the light blue dashed curve, partitions the network into $\lbrace1, 2, 3, 4, 5\rbrace$ and $\lbrace6, 7, 8, 9\rbrace$. The links in this cut are $(4,6)$ and $(5,6)$, each with capacity of 2. The total capacity of this cut is therefore $4$.  We call this the secondary bottleneck since this cut has the second lowest capacity: that cut with a capacity closest to the minimum cut capacity. 

There is a useful relationship between the maximum flow $F^*$ and the problem of finding the bottleneck.  This arises from the famous Max-flow min-cut theorem in graph theory~\cite{AhujaNetworkFlows}.  The maximum flow and the minimum cut hold a dual relationship, namely, the maximum amount of flow that can be transmitted through a given flow network is equal to the capacity of the minimum cut.   Now we define network cuts formally, and give a mathematical description of the Max-flow min-cut theorem.

\bigskip
\begin{addmargin}[1.5em]{1.5em}
For $S \subset V$, denote $\overline{S} := V - S$. The sets $S$ and $\overline{S}$ form a partition of $V$, and the set of arcs from $S$ to $\overline{S}$, denoted by
$$
\d^{+}(S) = \{(v,w) \in E: v \in S, w \in \overline{S}\},
$$
is called a \emph{cut} of $G$ induced by $S$.  If the source $s \in S$, then $\d^{+}(S)$ corresponds to the set of arcs leaving the set of nodes containing the source.

\bigskip
\begin{definition}
{\em A cut $\d^{+}(S)$ such that $s \in S$ and $t \in
\overline{S}$ is called an \emph{$(s,t)$-cut} of the flow network $(G,
u, s, t)$. The {\em capacity} of such a cut is defined to be
$$u(\d^{+}(S)) := \sum_{e \in \d^{+}(S)} u_e.$$}
\end{definition}
\end{addmargin}

\bigskip

\begin{addmargin}[1.5em]{1.5em}
\noindent\textbf{The Minimum Cut Problem (MCP):} Given a directed graph $G=(V,E)$ with two specific nodes,
$s$ (source) and $t$ (sink), and a capacity $u_e \ge 0$ on
each arc $e \in E$, find an $(s, t)$-cut $\d^+(S)$ with minimum
capacity $u(\d^+(S))$.

\bigskip
\noindent
The Minimum Cut Problem is equivalent to the MFP by the following theorem.

\begin{theorem}
\label{thm:min-max}{\bf Max-flow min-cut theorem:} In any network $(G,
u, s, t)$, the maximum value of a feasible $(s, t)$-flow is equal
to the minimum capacity of an $(s, t)$-cut. That is,
\bea
\max\{f(x): \mbox{\rm $x$ a feasible $(s, t)$-flow}\} = \min\{u(\d^{+}(S)): \mbox{\rm $\d^{+}(S)$ an $(s, t)$-cut}\}. \nonumber
\eea
\end{theorem}
\end{addmargin}

It is easy to see that the minimum cut set $B$ \emph{is a subset} of $\mathcal{P}$. Forward arcs in the minimum cut must be saturated (each arc has flow value equal to its capacity). If any arc in the minimum cut is not saturated, then $F^*$ is not the maximum flow as the flow can be further increased.  As mentioned, all the above optimization problems can be solved efficiently, and a number of polynomial time algorithms for them have been discovered, as discussed in~\cite{AhujaNetworkFlows}. For the MCFP, we used a version of the Ford-Fulkerson algorithm called the Edmonds-Karp Algorithm. The Ford-Fulkerson is relatively simple to describe. The algorithm keeps looking for a path from the source to the sink which has unused capacity on every link (such paths are called augmenting paths). It then increases the flow on this path and repeats this process until there are no more augmenting paths. The complexity of the Edmonds-Karp algorithm is $O(nm^2)$, where $n$ is the number of nodes and $m$ is the number of arcs. This is a polynomial complexity and is therefore considered efficient.

\subsection{Measures of robustness from $\mathcal{P}$, $B$ and $B_{min}$}
\label{secRob}

We propose two measures of transmission robustness: $p_{min}$ to quantify the pathway redundancy in $\mathcal{N}$, and a reroute score $R_{re}$ to quantify the extent to which the available redundancy is exploited to redistribute forces after link failure.

\subsubsection{Pathway redundancy $p_{min}$}
A measure of robustness must necessarily consider the pathway redundancy in $\mathcal{N}$. Pathway redundancy focusses on the multiplicity of paths from source to sink in $G$ with respect to the topology of the uncapacitated directed links of $\mathcal{N}$. Transmission robustness, in analogy to maximum flow, has a useful duality property: (1) the maximum number of percolating link-disjoint paths (paths that do not overlap or share links) through $\mathcal{N}$ between the source and the sink is equal to (2) the minimum number of links whose removal would disconnect the source and the sink from each other. We refer to (1) as the \textit{pathway redundancy} $p_{min}$ of $G$. The dual quantity (2) is referred to as the \textit{link-connectivity} of $G$, and is equal to the minimum number of links in any cut separating the source and the sink. The duality of (1) and (2) is a consequence of the Max-flow min-cut theorem, often referred to as Menger's Theorem~\cite{AhujaNetworkFlows} when dealing with uncapacitated networks such as ${G}$.

A cut which separates the source and the sink, and has the minimum number of links, will be denoted by $B_\mathrm{min}$. Note that in the example of Figure \ref{fig:mf}, a minimum of two links must be removed in order to disconnect $s$ from $t$. Therefore $p_{min}=2$: that is, there are two available link-disjoint paths from source to sink.  From the dual perspective, the maximum number of link-disjoint paths between $s$ and $t$ is also $2$. This can be seen from the fact that every path from $s$ to $t$ must use one of the two links $(4,6)$ or $(5,6)$. There are a number of candidates for $B_\mathrm{min}$ in this example, including the secondary bottleneck. Other possible candidates include the pairs $(6,8), (6,7)$ and $(6,8), (7,9)$.

The pathway redundancy $p_{min}$ can be calculated algorithmically as follows: first we assign a capacity of $u'_e:=1$ to each link $e$ of $G$ and let the resultant flow network be denoted by $\mathcal{F}_1=(G,u',s,t)$. We then calculate the maximum flow $F_1$ from $s$ to $t$ in $\mathcal{F}_1$ and let $B_\mathrm{min}$ be a minimum cut of $\mathcal{F}_1$. Note that since each link is of unit capacity in $\mathcal{F}_1$, the magnitude of $F_1$ is equal to the number of links in the cut $B_\mathrm{min}$. Thus $p_{min}=F_1$.

Two aspects need further clarification. First, since each link is of unit capacity in $\mathcal{F}_1$, each unit of flow that is sent from $s$ to $t$ must use a unique path. This is consistent with the fact that the pathway redundancy of $G$ is equal to the maximum number of link-disjoint paths from $s$ to $t$. Each such path corresponds to (and uses) a unique link in $B_\mathrm{min}$. Therefore, removing the links of $B_\mathrm{min}$ from $G$ will disconnect $s$ from $t$, and no smaller set of links, when removed from $G$, will disconnect $s$ from $t$. Second, since $B_\mathrm{min}$ has the minimum number of links of any cut, the cardinality of $B_\mathrm{min}$ is a lower bound on the cardinality of $B$, the minimum cut of our original flow network $\mathcal{F}$.

\subsubsection{Reroute score $R_{re}$}
When a contact between two grains breaks, certain paths will no longer be available for force transmission. If there is high pathway redundancy in the network then flow can be diverted without significantly affecting the maximum flow $F^*$. This capability to reroute flow can be quantified through a {\it reroute score}, denoted by $R_{re}$, given by
\begin{equation}
R_{re} = \rho\left( 1 - \frac{|\alpha - \gamma|}{(\alpha + \gamma + 1)} \right),
\label{eqn:reroute}
\end{equation}
here $0 \leq \rho \leq 1$ where $\rho$ is the ratio of the number of links in $\mathcal{P}$ relative to its initial value prior to damage, $\alpha \geq 0$ is the number of replacement links to which flow is diverted, and $\gamma \geq 0$ is the number of links that cease to be part of $\mathcal{P}$ (links that leave $\mathcal{P}$ either to join the complementary set $\overline{\mathcal{P}}$ in $\mathcal{N}$ or are damaged).  The reroute score is maximum when a replacement contact from $\mathcal{N}$ is found for every contact lost to $\mathcal{P}$: $R_{re}=\rho$ if $\alpha = \gamma$, where $\rho=1$ when there is no damage to $\mathcal{N}$.



\section{Results}
\label{sec:results}

Most of the key trends in the evolution of force transmission and fracture patterns in Data I and II are qualitatively similar; for these cases, we mainly focus the summaries on Data I for brevity.  A discussion of both data sets is presented where important differences exist.

\subsection{Quantifying transmission strength from $F^*$}

\begin{figure}[H]
\centering
\begin{picture}(420,150)(0,0)
  \put(5,0){\includegraphics[width=.47\textwidth,clip]{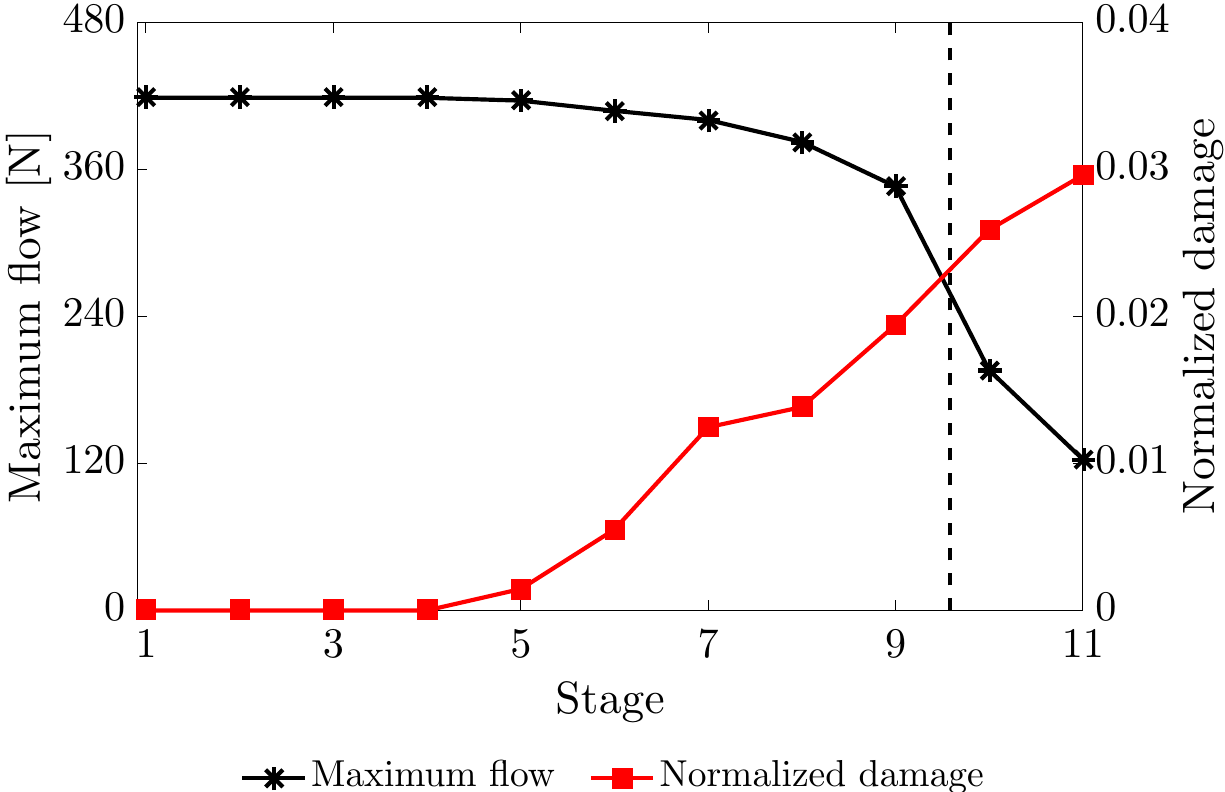}}
  \put(220,0){\includegraphics[width=.47\textwidth,clip]{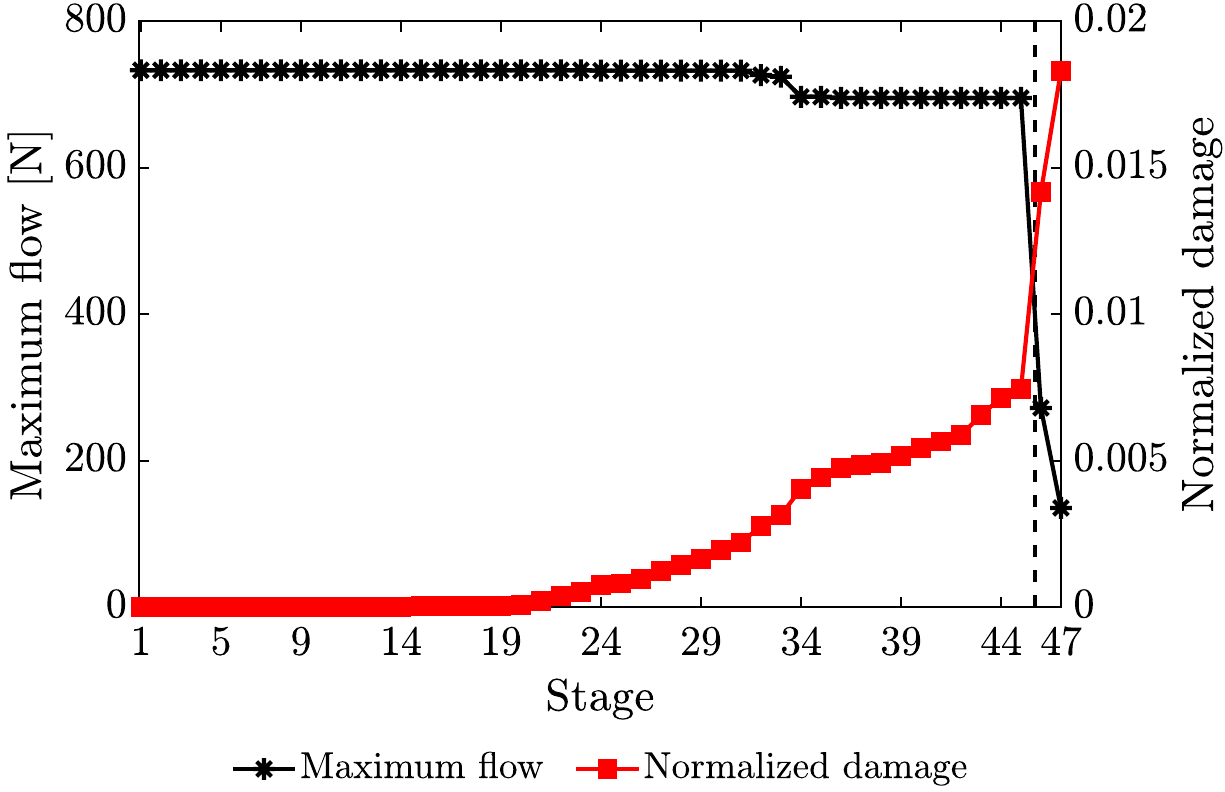}}
  \put(-5,140){\small{(a)}}
  \put(215,140){\small{(b)}}
\end{picture}
\caption{(Color online) {\it Maximum flow quantifies the global transmission capacity of the bonded contact network.}  Evolution of the maximum flow $F^*$ and the normalized damage (number of broken bonded contacts relative to the initial number of bonded contacts) in the bonded contact network $\mathcal{N}$ for: (a) Data I and (b) Data II. Dashed vertical line marks the stage at peak load. }
\label{fig:mfnd}      
\end{figure}

As damage initiates and propagates, we observe a steady decline in the global transmission capacity of both specimens as measured by the maximum flow $F^*$ (Figure~\ref{fig:mfnd}).  This trend highlights an underlying degradation in force transmission in $\mathcal{N}$, and of the robustness of $\mathcal{N}$,  consistent with the progressive spread of damage.  The highest drop in $F^*$ coincides with the transition from pre-failure to the failure regime: stages 9-10 for Data I and stages 45-46 for Data II, as seen in the evolution of the macroscopic vertical tensile load (Figure~\ref{fig:bb}).  


\subsection{Predicting tensile force chains from $\mathcal{P}$}

\begin{figure}[H]
\centering
\begin{picture}(420,270)(0,0)
  \put(0,140){\includegraphics[width=1\textwidth,clip]{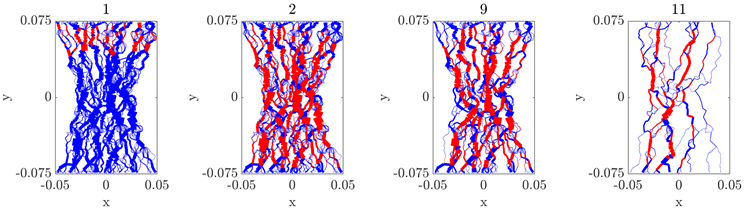}}
  \put(-10,5){\includegraphics[width=.25\textwidth,clip]{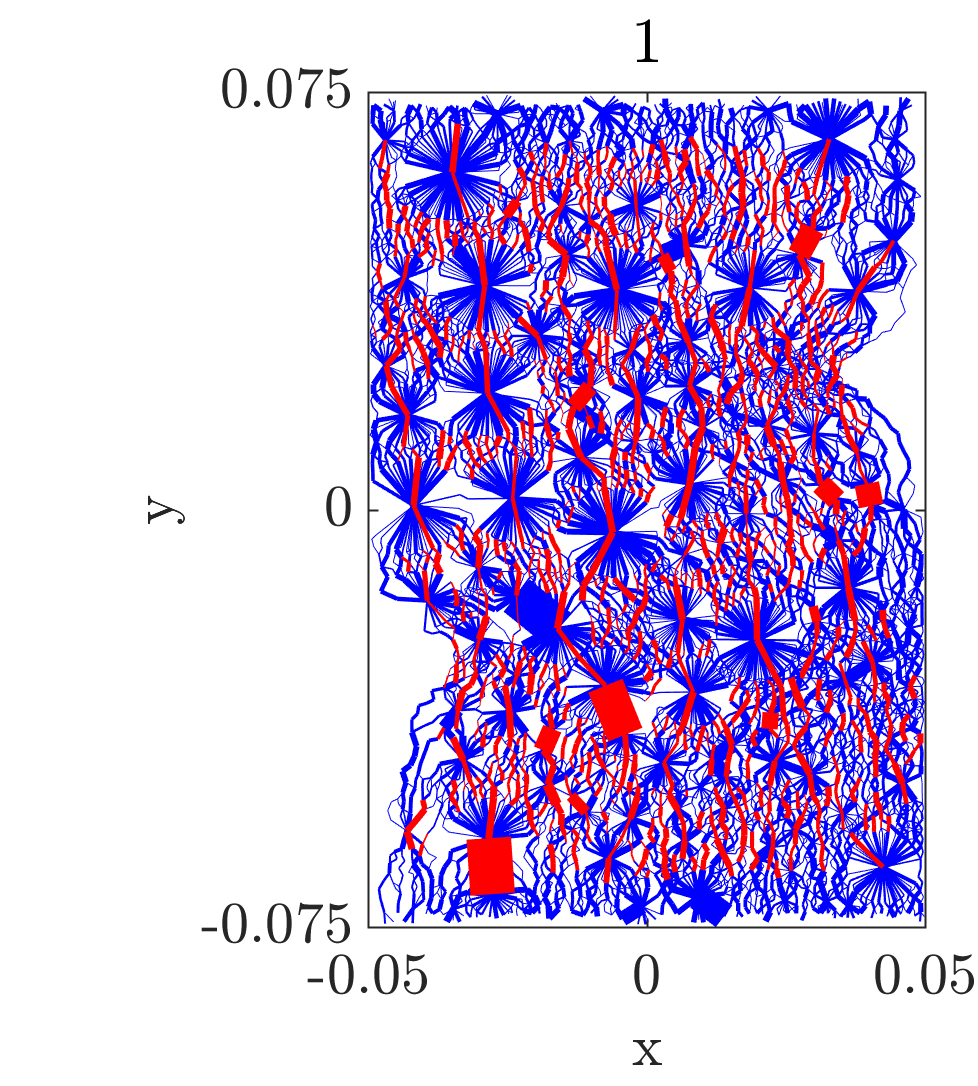}}
  \put(100,5){\includegraphics[width=.25\textwidth,clip]{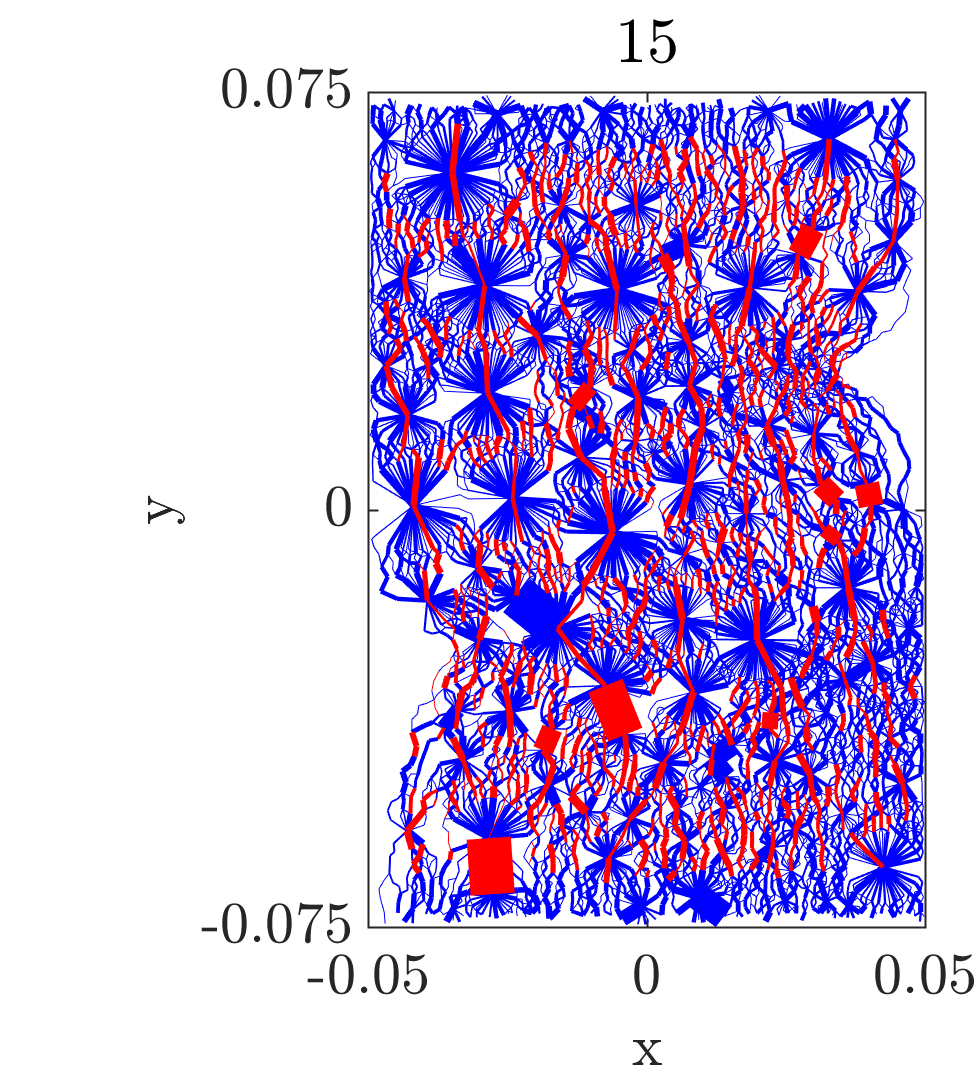}}
  \put(210,5){\includegraphics[width=.25\textwidth,clip]{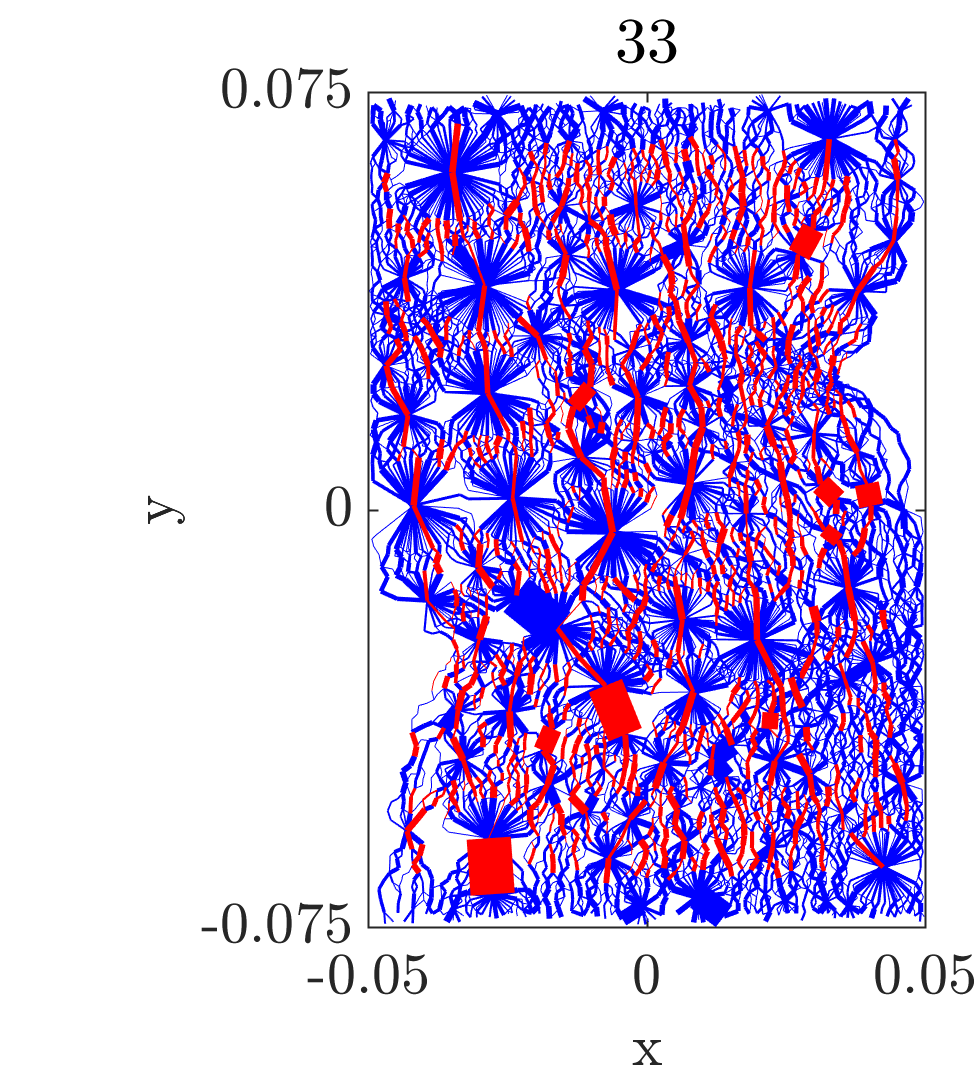}}
  \put(320,5){\includegraphics[width=.25\textwidth,clip]{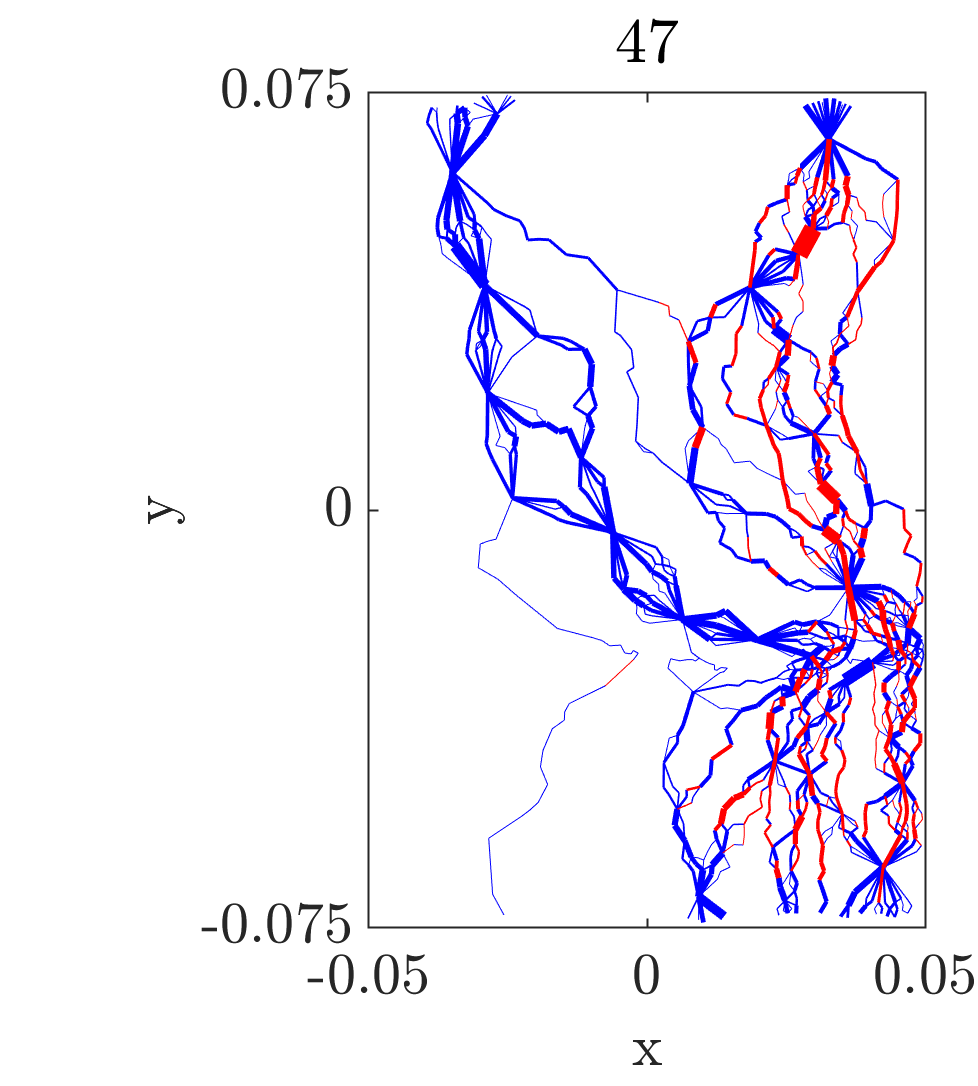}}
  \put(-5,260){\small{(a)}}
  \put(-5,120){\small{(b)}}
\end{picture}
\caption{(Color online) {\it Most tensile force chains lie along the shortest possible routes that transmit the global capacity in the direction of the applied tensile load.}  Optimized routes $\mathcal{P} \in \mathcal{N}$ are the interconnected blue and red links that form percolating transmission pathways through $\mathcal{N}$ for: (a) Data I and (b) Data II. Link thickness is proportional to the force flow across each link.  Links between grains in tensile force chains are colored red; otherwise the link is colored blue. Residual strength due to $\mathcal{P}$ links along the left (right) side of the specimen in Data I (Data II).  Note percolating tensile force chains are first established at stage 2 in Data I (Figure 1, Section 1 of \cite{DiB2018}).   }
\label{fig:ptfcs}      
\end{figure}

The optimized routes $\mathcal{P}$ which transmits $F^*$ along the shortest possible paths are shown in Figure~\ref{fig:ptfcs}.  On average, $\mathcal{P}$ comprises around $21\%$ of $\mathcal{N}$ links before peak and around $5\%$ after peak load for Data I (around $46\%$ of $\mathcal{N}$ links before peak and around $7\%$ after peak load for Data II).  We find a potential for $\mathcal{P}$ to predict the preferential paths for tensile force transmission, namely, the tensile force chains (Section 1 of \cite{DiB2018}).  The spatio-temporal evolution of tensile force chains in $\mathcal{P}$ and the composition of $\mathcal{P}$, as shown in Figure~\ref{fig:ptfcs} and Sections 2-3 of \cite{DiB2018}, highlights three salient trends common to Data I and II. First, the majority of tensile force chain contacts are in $\mathcal{P}$ in the pre-failure regime: $73\%$ across stages 2-9 for Data I, $91\%$ across stages 1-45 for Data II.  Second, while scalar force flows cannot generally be used to predict the magnitudes of the vectorial contact forces, the direction of the normal tensile contact force anisotropy of grains in tensile force chains (vertical, as shown in Figure~\ref{fig:consRule} (b)) is consistent with the {\it orientation} of the preferred paths for flow.  In this study, we do not use the force flows except as a visual guide in Figure~\ref{fig:ptfcs} where they serve solely to highlight the orientation of the preferential paths for force flow.  It can be seen that the thickest links (highest force flows) generally align with the vertical direction --- as expected since the Minimum Cost Flow Problem (MCFP) “pushes” as much flow as possible through the most direct paths (i.e., paths with least cost or fewest member links) from the source to the sink.  Thus those contacts or links in alignment with the direction of flow (vertical for our samples) are favored over those aligned in the transverse direction, which are either not used (links not in $\mathcal{P}$ with zero flow) or have smaller force flow values.  Indeed most links in $\mathcal{P}$ are strong contacts, i.e., transmit force above the global mean tensile force magnitude. Third, the sudden breakdown of pathways in $\mathcal{P}$ coincides in space and time with the collective failure of tensile force chains in stages 10-11: note the marked loss of $\mathcal{P}$ links along the right side of the specimen for Data I (left side of the specimen for Data II).  Thus the residual strength is mainly due to the flow pathways in $\mathcal{P}$ along the left (right) side of the specimen in Data I (Data II).

Relative to the specimen in Data I, the much higher ratio of fine cement matrix to coarse aggregate grains, with some aggregates being bigger and some cement grains being smaller in the specimen in Data II --- lead to one important difference between the optimized routes $\mathcal{P}$ in the two specimens: a relative abundance of supporting contacts (blue) on either side of tensile force chain contacts (red) in the pre-failure regime in Data II, as shown in Figure~\ref{fig:ptfcs} and in Section 3 of \cite{DiB2018}.  Since there are many more supporting contacts that can share and take the load off, or replace damaged, tensile force chain contacts in Data II, this specimen is stronger in tension and more resistant to damage compared to the specimen in Data I.

\subsection{Predicting the ultimate crack pattern and crack interaction from force bottlenecks $B$}

\begin{figure}[H]
\centering
\begin{picture}(420,390)(0,0)
  \put(0,0){\includegraphics[width=.95\textwidth,clip]{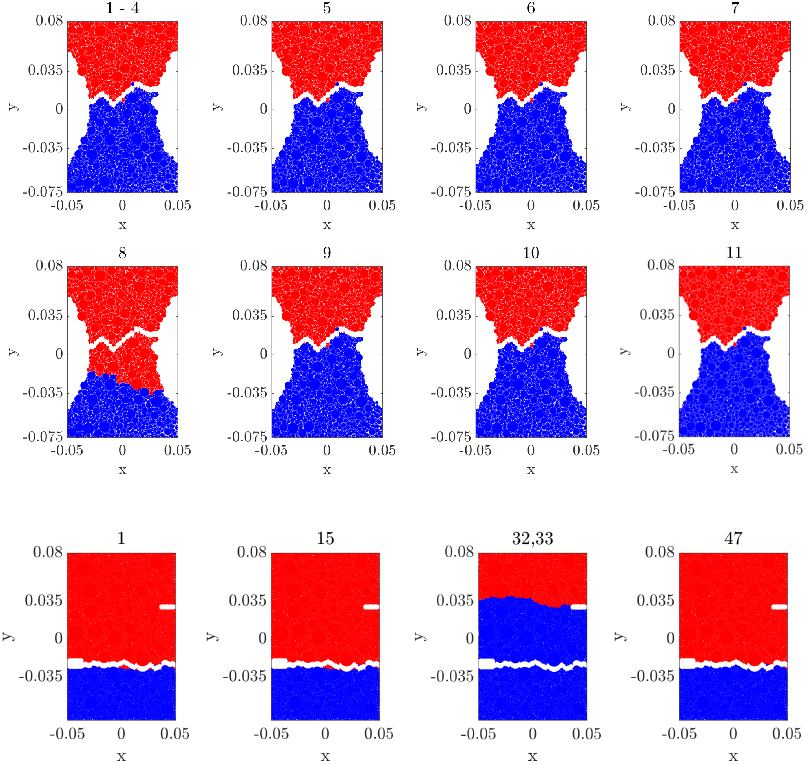}}
  \put(-5,385){\small{(a)}}
  \put(-5,120){\small{(b)}}
\end{picture}
\caption{(Color online) {\it Force bottlenecks provide an early prediction of the ultimate crack pattern.} Spatial distribution of the bottlenecks $B$ (red-blue interface) for: (a) Data I and (b) Data II. Grains on either side of the ultimate macrocrack are artificially separated to aid visual comparison of the bottleneck versus the actual macrocrack location.  Note a recurring bottleneck $B=B^{*}$ persists for all stages except at stage 8 (stages 32 and 33) when a second bottleneck $B=B_{8}$ emerges for Data I ($B=B_{32}=B_{33}$ for Data II).}
\label{fig:minCut}      
\end{figure}

The force bottlenecks $B$, identified using the minimum cut, are shown in Figure~\ref{fig:minCut}.  Bottlenecks in the pre-failure regime form in two separate locations of the specimen.  In Data I, the primary bottleneck $B^{*}$ persists in the middle of the gauge region or neck of the specimen across all stages except stage 8 (Figure~\ref{fig:minCut} (a)).  This recurring bottleneck predicts the location of the dominant macrocrack: the primary crack that forms during failure (stages $10-11$).  The second bottleneck $B_{8}$ emerges momentarily in stage $8$ in the lower section of the specimen.  $B_{8}$ predicts the secondary crack. Similar trends apply in Data II: the primary bottleneck $B^{*}$ persists for stages 1-31 (lower section), while the secondary bottleneck $B_{32}$ forms at stages 32-33 (upper section).

\begin{figure}[H]
\centering
\begin{picture}(420,310)(0,0)
  \put(0,160){\includegraphics[width=.45\textwidth,clip]{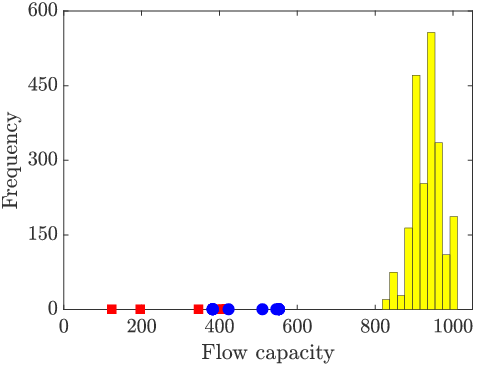}}
  \put(210,160){\includegraphics[width=.45\textwidth,clip]{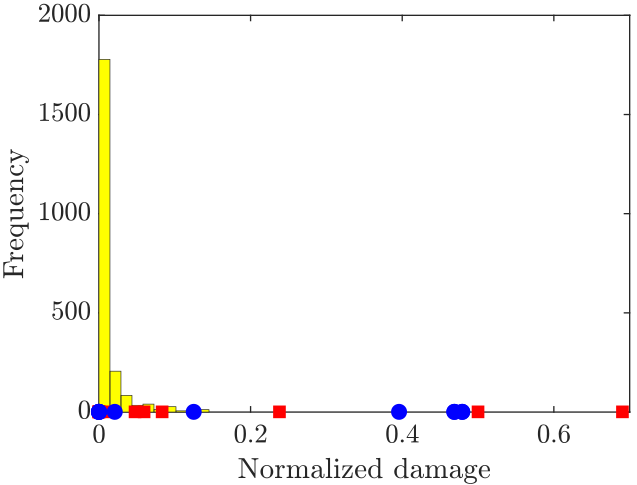}}
  \put(250,220){\includegraphics[width=.34\textwidth,clip]{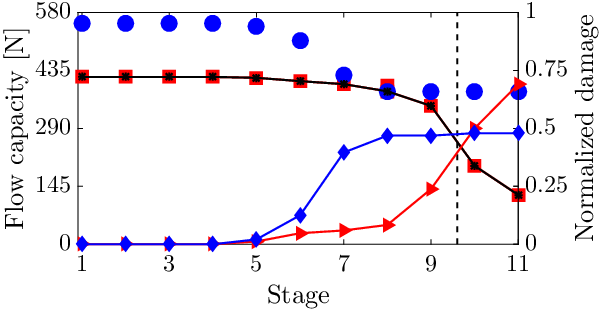}}
  \put(270,208){\includegraphics[width=.28\textwidth,clip]{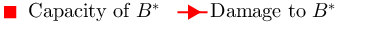}}
  \put(34,280){\includegraphics[width=.15\textwidth,clip]{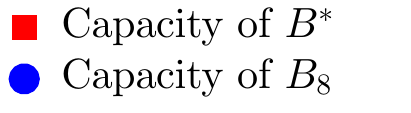}}
  \put(270,200){\includegraphics[width=.28\textwidth,clip]{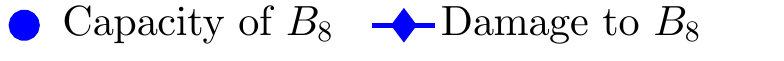}}
  \put(268,192){\includegraphics[width=.14\textwidth,clip]{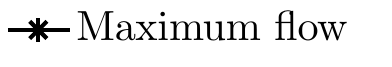}}   
  \put(0,0){\includegraphics[width=.45\textwidth,clip]{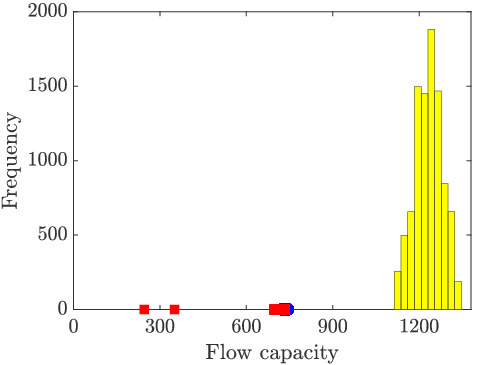}}
  \put(210,0){\includegraphics[width=.45\textwidth,clip]{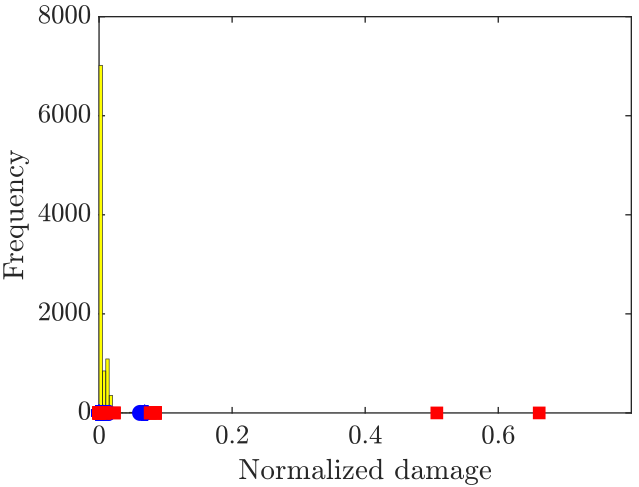}}
  \put(255,65){\includegraphics[width=.32\textwidth,clip]{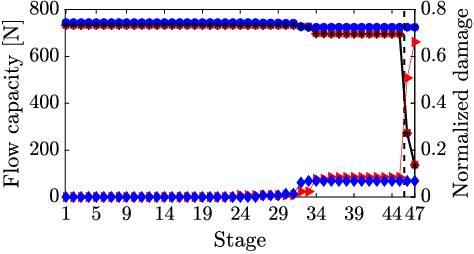}}
  \put(284,86){\includegraphics[width=.13\textwidth,clip]{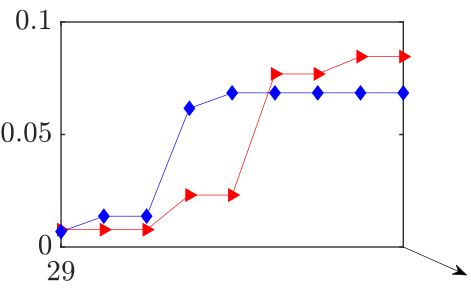}}
  \put(34,117){\includegraphics[width=.18\textwidth,clip]{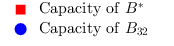}}
  \put(269,55){\includegraphics[width=.28\textwidth,clip]{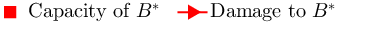}}
  \put(269,45){\includegraphics[width=.3\textwidth,clip]{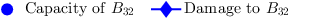}}
  \put(266,35){\includegraphics[width=.14\textwidth,clip]{figuresManuscript/capDam6.png}}
  \put(-5,305){\small{(a)}}
  \put(205,305){\small{(b)}}
  \put(-5,150){\small{(c)}}
  \put(205,150){\small{(d)}}
\end{picture}
\caption{(Color online) {\it Cut capacity distinguishes the bottlenecks and their damage evolution.}  (a,b) Data I and (c,d) Data II. Frequency distribution for (a,c) the capacities of $200$ random cuts of $\mathcal{N}$, and (b,d) their normalized damage across {\it all} observed stages of loading.  The capacities of the primary bottleneck $B^{*}$ (secondary bottleneck $B_8$ for Data I or $B_{32}$ for Data II) are marked by the red squares (blue circles).  Normalized damage in a given bottleneck is the number of broken bonded contacts relative to the number of bonded contacts in the initial stage 1.  Damage in bottlenecks prior to stage 30 is confined to one broken bond at: stage 24 in $B^{*}$ and stage 27 in $B_{32}$. Inset shows crack interaction from the perspective of the co-evolution of the capacities and damage in the bottlenecks: dashed vertical line marks the stage at peak load.} 
\label{fig:arb}       
\end{figure}

To understand why the minimum cut provides an {\it early prediction of the location of the ultimate crack pattern}, it is essential to examine the capacity of this cut relative to those of arbitrary cuts that partition $\mathcal{N}$.  As shown in Figure~\ref{fig:arb} (a,b) for Data I, the cut capacity distinguishes $B^{*}$ and $B_{8}$, the two disjoint force bottlenecks\footnote{$B^{*}$ and $B_{8}$ do not have common links and thus their capacities are entirely independent of each other.}, from other cuts even before damage: the capacities of $B^{*}$ and $B_{8}$ are significantly less than those of other cuts for all of loading history (Figure~\ref{fig:arb} (a)).  This explains why the bottlenecks are in turn highly prone to congestion and why they manifest a distinct pattern of damage evolution relative to those of other cuts (Figure~\ref{fig:arb} (b)).  Specifically, while damage values sustained by $B^{*}$ and $B_{8}$ are comparable to other cuts in the initial stages of the pre-failure regime, these rapidly increase close to peak load (stage 9) and consistently exceed those of other cuts of $\mathcal{N}$ during failure (stages 10-11).   The greatest damage occurs initially in the secondary bottleneck $B_{8}$ up until stage 9, after which damage concentrates in the primary bottleneck $B^{*}$ (Figure~\ref{fig:arb} (b) inset) where the macrocrack ultimately forms (recall Figure~\ref{fig:minCut}). Figure~\ref{fig:arb} (c,d) for Data II shows the same trends.  More damage concentrates in the second bottleneck $B_{32}$ across stages 30-34 (note only 1 bond breaks in $B^{*}$ and in $B_{32}$ prior to stage 30).  But as the tensile load rises, damage in $B^{*}$ also rises, eventually matching that of $B_{32}$ across stages 35-36, before the abrupt cascade of bond failures across stages 45-47. Thus the bottlenecks identify the most vulnerable as well as the most critical sites of the specimen, keeping in mind the bottleneck capacity sets the maximum flow $F^*$ (i.e., the upper bound on the force flow that can be transmitted through $\mathcal{N}$), as discussed earlier in Section \ref{sec:methods}.

The evolution of the distributions in cut capacities and attendant damage in Figure~\ref{fig:arb} also provides new insight into crack interaction in the presence of damage.   With both distributions in mind, and the fact that the global transmission strength $F^*$ is equal to the minimum cut capacity, consider now two possible scenarios at some equilibrium state early in the pre-failure regime for the specimen in Data I. Scenario S1 is where the capacity of the minimum cut is much smaller than any other cut (e.g., $B^{*}$ in stages 1-6, Figure~\ref{fig:arb} (a)).  Scenario S2 is where multiple potential force bottlenecks exist, meaning one or more cuts have capacities that are close to the minimum value (e.g., $B_{8}$ in stages 7-9, Figure~\ref{fig:arb} (b) inset).  In scenario S1, a single recurring bottleneck would likely emerge in the pre-failure regime (stages 1-6, Figure~\ref{fig:minCut} (a)), since a few links can be lost to other cuts without change to the minimum cut capacity (stage 5-6, Figure~\ref{fig:arb} (b)).  By contrast, in scenario S2, we can expect changes to the bottlenecks in the pre-failure regime (stages 7-9, Figures~\ref{fig:minCut} (a) and \ref{fig:arb} (a,b)), since the loss of one to a few links in another cut with only a slightly higher capacity would suffice to reduce its capacity to the global minimum.  In turn, this cut would become the new minimum cut (bottleneck).  This is seen in the switches in bottleneck: from $B^{*}$ to $B_{8}$ across stages 7-8, and vice versa across stages 8-9.  This process whereby another cut (a secondary force bottleneck) takes the ``fall'' by loosing its links in place of the previous minimum cut sustaining further loss of links (a condition that would guarantee a reduction in global transmission strength $F^*$) --- ensures that the inevitable reduction in global transmission strength $F^*$ is either delayed or at least minimized, as damage spreads.  Qualitatively similar trends manifest between $B^{*}$ and $B_{32}$ in Data II.  In summary, results here suggest that {\bf multiple crack interaction in the pre-failure regime can be viewed as a cooperative process among force bottlenecks to minimize the inevitable reduction in global force transmission capacity due to damage.}

\subsection{Quantifying robustness from $\mathcal{P}$, $B$ and $B_\mathrm{min}$ }

Redundancy in transmission pathways underpins system robustness to damage.  Here we quantify system robustness as topological connectivity evolves using a measure of the redundancy in force pathways and a measure of the extent to which this redundancy is exploited to redistribute forces after successive link or contact failures occur.

\subsubsection{Pathway redundancy $p_{min}$}

\begin{figure}[H]
\centering
\begin{picture}(420,550)(0,0)
 \put(5,410){\includegraphics[width=1\textwidth,clip]{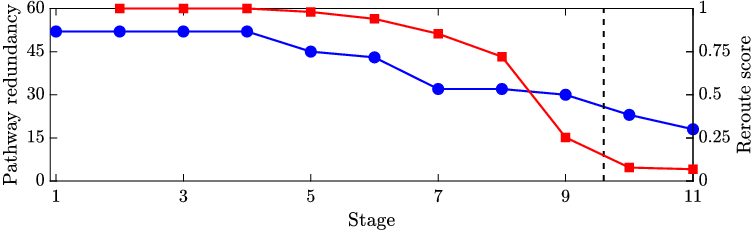}}
 \put(40,442){\includegraphics[width=.33\textwidth,clip]{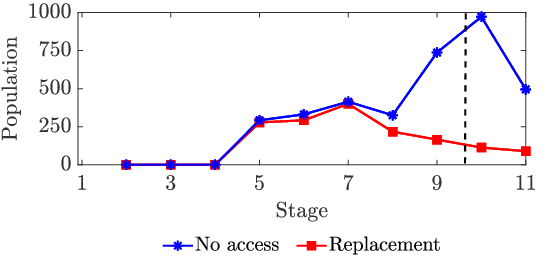}}
 \put(5,245){\includegraphics[width=1\textwidth,clip]{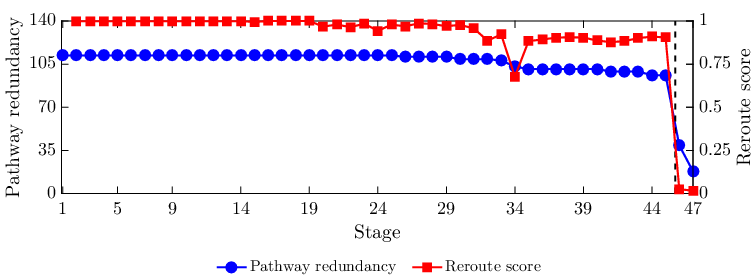}}
 \put(50,296){\includegraphics[width=.35\textwidth,clip]{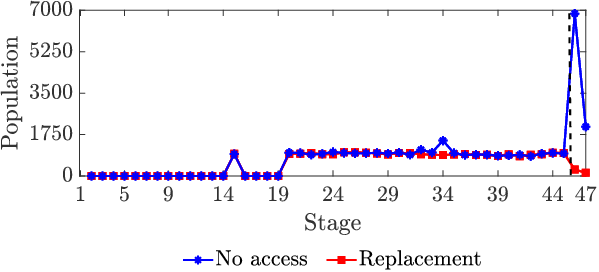}}
 \put(50,135){\includegraphics[width=.3\textwidth,clip]{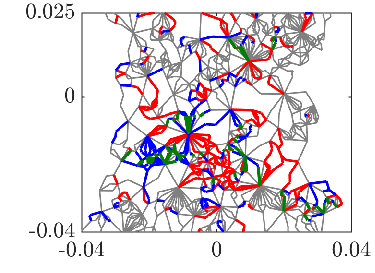}}
 \put(245,118){\includegraphics[width=.3\textwidth,clip]{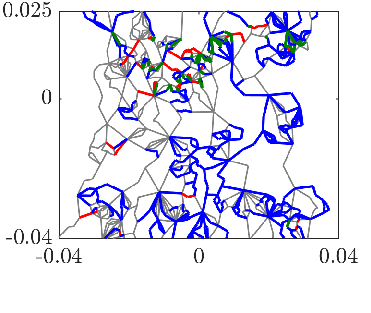}}
 \put(57,25){\includegraphics[width=.3\textwidth,clip]{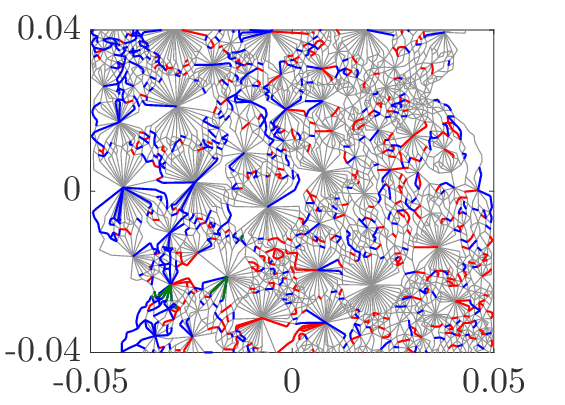}}
 \put(245,25){\includegraphics[width=.32\textwidth,clip]{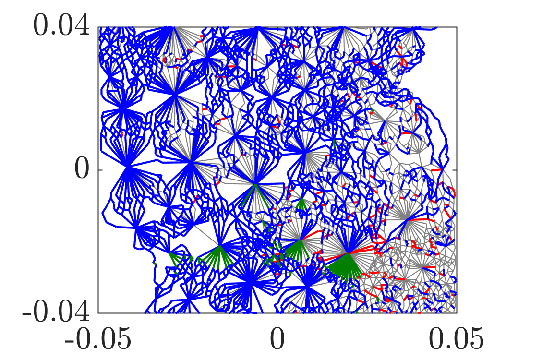}}
 \put(80,5){\includegraphics[width=.35\textwidth,clip]{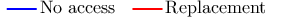}}
 \put(215,5){\includegraphics[width=.35\textwidth,clip]{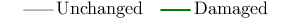}}
\put(0,550){\small{(a)}}
\put(0,400){\small{(b)}}
\put(0,240){\small{(c)}}
\put(0,120){\small{(d)}}
\put(119,226){\small{6-7}}
\put(305,225){\small{9-10}}
\put(119,112){\small{33-34}}
\put(305,112){\small{45-46}}
\end{picture}
\vspace{-.25cm}
\caption{(Color online) {\it Pathway redundancy, which allows rerouting of forces to alternative paths, underpins robustness of $\mathcal{N}$.} (a,c) Data I and (b,d) Data II.  (a,b) Evolution of the pathway redundancy $p_{min}$ and the reroute score $R_{re}$ ($R_{re}=1$ is maximum when a new replacement link from $\mathcal{N}$ is found for every old link that is no longer accessed in $\mathcal{P}$) for Data I.  Inset shows the evolution of the number of links that leave $\mathcal{P}$ but remain in $\mathcal{N}$ (no access) and enter $\mathcal{P}$ (replacement) due to rerouting.  Dashed vertical line marks the stage at peak load. (c,d) Spatial distribution of rerouted and damaged links in $\mathcal{P}$.  }.
\label{fig:rerouted}       
\end{figure}
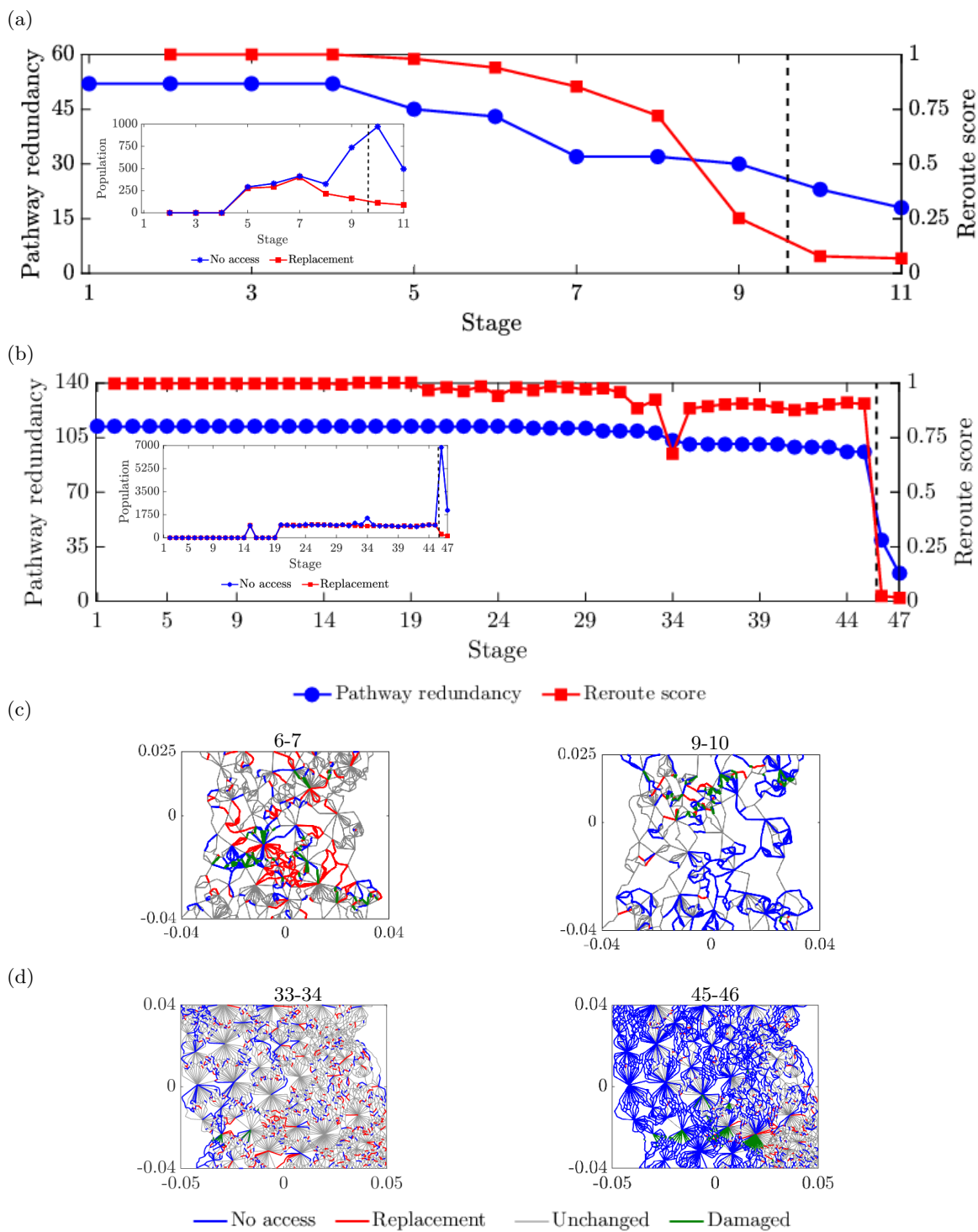

Multiple pathways are available for the transmission of the applied tensile load. Figure~\ref{fig:rerouted} (a,b) shows the evolution of  $p_{min}$, the number of available transmission pathways that do not share links (non-overlapping paths) between the top and bottom walls of the specimen.  The monotonic decrease in $p_{min}$ is consistent with the progressive spread of damage in $\mathcal{N}$ for both specimens. The higher pathway redundancy for Data II supports the earlier finding that tensile force chain contacts in $\mathcal{P}$ are significantly more supported than those in Data I (recall Figure~\ref{fig:ptfcs}).


\subsubsection{Reroute score $R_{re}$}

The system exploits pathway redundancy in the face of damage by continually reconfiguring: the force pathways, and the distribution of forces transmitted through them.  Here we quantify the former, the process of rerouting forces to alternative pathways, with respect to the evolution of links in $\mathcal{P}$ and $\mathcal{N}$ using the reroute score $R_{re}$.  At the onset of damage in $\mathcal{N}$, which is at stage 5 for Data I and at stage 15 for Data II, old links leave as new links enter $\mathcal{P}$ (Section 4 of \cite{DiB2018}).  An old link that ceases to be part of $\mathcal{P}$ can be either damaged (link at stage $t-1$ breaks and no longer exist in $t$) or the link is a member of the complement set of $\mathcal{P}$ (link is no longer accessed in $\mathcal{P}$ but exists in $\mathcal{N}$).  The evolution of $R_{re}$ accurately tracks the transition from pre-failure to post-failure regime (Figure~\ref{fig:rerouted} (a,b)).

For Data I, across stages 1-8, the system compensates for the disruptions to transmission by finding replacement contacts in $\mathcal{N}$ for almost every contact that is no longer accessed in $\mathcal{P}$.  By contrast, across stages 9-11, we observe a sudden degradation in $\mathcal{P}$ without a matching recovery: note the surge in the number of contacts that leave $\mathcal{P}$ at the same time as feasible replacement contacts from $\mathcal{N}$ dwindle in numbers, as shown in Figure~\ref{fig:rerouted} (c) and Section 4 of \cite{DiB2018}).  Force rerouting patterns corroborate the cooperative behavior in the bottlenecks in the early stages of the pre-failure regime, as discussed previously in Section 5.3.  $B_8$ sustains most of the damage and, consequently, force reroutes up until stage 8 (Section 4 of \cite{DiB2018}); this leaves $B^*$ essentially intact with minimal damage and change to its tensile force chain membership across stages 1-8 (Figure~\ref{fig:evolTT} (a)).

For Data II, across stages 1-44, the high level of redundancy in $\mathcal{N}$ enable significant force redistributions (i.e., changes in the force magnitudes) across the different contacts in $\mathcal{P}$ with minimal bond breakage.  This and the significant number of alternative pathways surrounding tensile force chains in $\mathcal{P}$, as shown in Figure~\ref{fig:ptfcs} and Section 3 of \cite{DiB2018} result in: (i) minimal damage to the bottlenecks as damage occurs elsewhere in the sample (only one damaged link in each until stage 30), and (ii) a weaker interaction among the bottlenecks.  Across stages 30-34, $B_{32}$ sustains greater damage and number of undamaged links that are lost to $\mathcal{P}$ (no access links,  Section 4 of \cite{DiB2018}).  This leaves $B^*$ essentially intact with minimal damage and change to its tensile force chain membership across stages 1-35 (Figure~\ref{fig:evolTT} (b)).  Across stages 36-45, damage is diverted away from both bottlenecks due to the high redundancy, before a cascade of bond breakages ensue in $B^*$ across stages 45-47.

\subsection{Brittle failure: suppressed followed by cascading failure in the force bottleneck} 

\begin{figure}[H]
\centering
\begin{picture}(420,420)(0,0)
	 \put(0,215){\includegraphics[width=1\textwidth,clip]{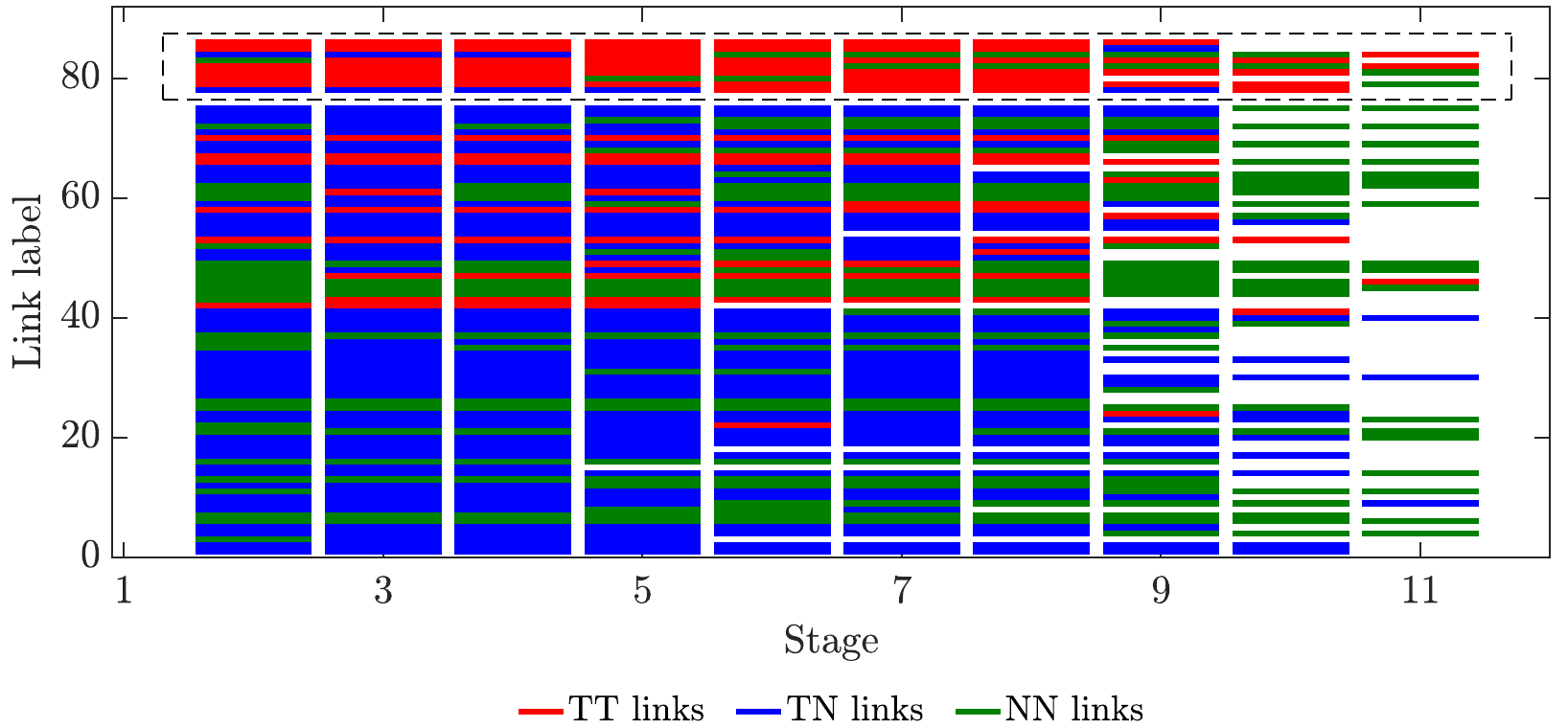}}
	\put(0,0){\includegraphics[width=1\textwidth,clip]{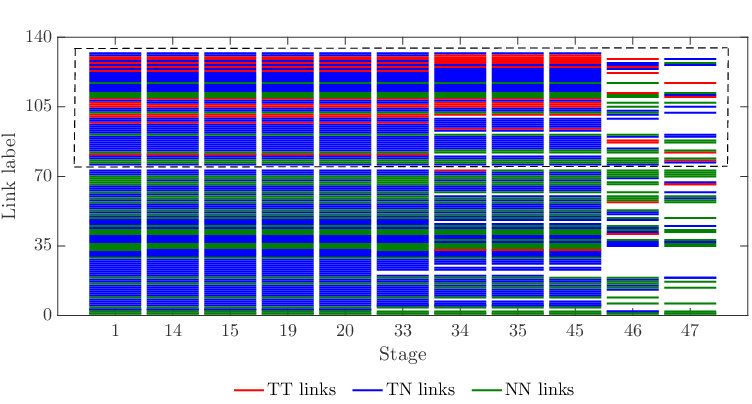}}
	\put(0,415){\small{(a)}}
	\put(0,215){\small{(b)}}
\end{picture}
\caption{(Color online) {\it Despite being prone to congestion, tensile force chains endure in the bottleneck due to damage being confined to low capacity links in the pre-failure regime. }  The evolution of link type for each of the $k$ member links of $B^*$ is shown for (a) Data I where $k=84$, and (b) Data II where $k= 130$.  Links in $B^*$ are ranked from lowest (1) to highest tensile bond capacity ($k$) at that stage when the tensile force chain network is first established (stage 2 for Data I and stage 1 for Data II).  Link type is represented by a horizontal bar colored according to the type of grains in contact (Sections 6.1-6.2 of \cite{DiB2018}): red (TT- tensile force chain grains), green (NN- neither is a tensile force chain grain), blue (TN- one is a tensile force chain grain, the other is not).  Dashed rectangle highlights links with above the global mean capacity.  A transition to a different link type manifests as a change in the color of the bar.  No bar is shown for a link that breaks.  }
\label{fig:evolTT}      
\end{figure}

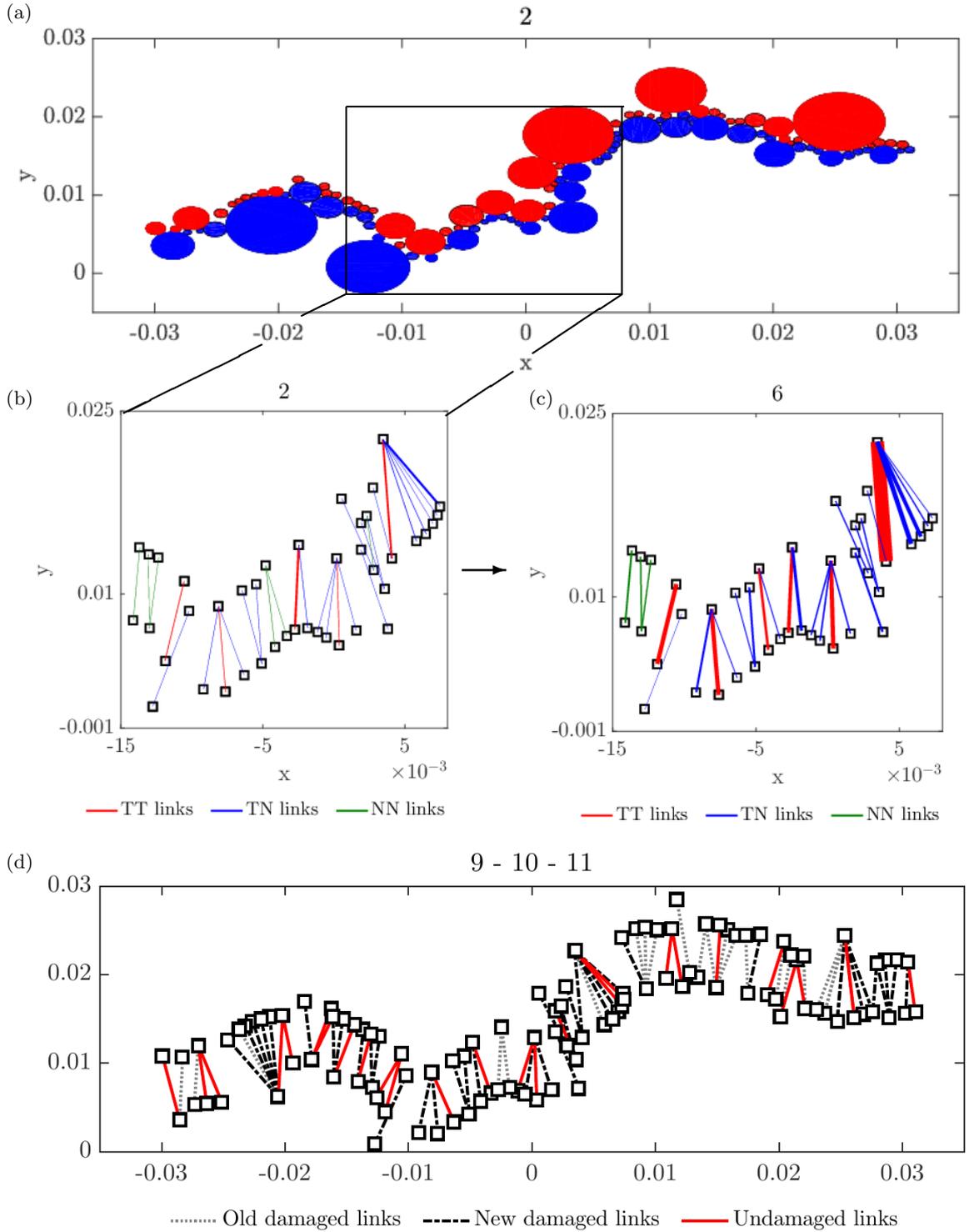
\begin{figure}[H]
\centering
\begin{picture}(420,540)(0,0)
  \put(0,390){\includegraphics[width=1\textwidth,clip]{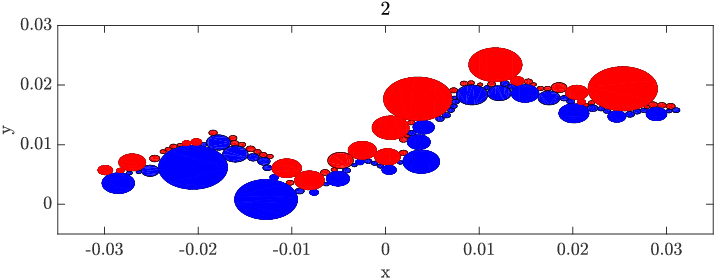}}
  \thicklines
  \put(148,510){\line(0,-1){85}}
  \put(272,510){\line(0,-1){85}}
    \put(148,510){\line(1,0){125}}
  \put(148,425){\line(1,0){124}}
    \put(6,187){\includegraphics[width=.45\textwidth,clip]{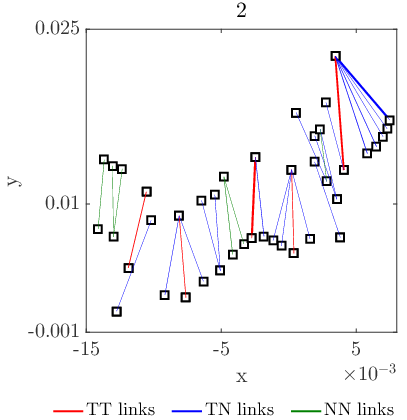}}
    \linethickness{1pt}
    \put(200,300){\vector(1,0){20}}
    \put(228,187){\includegraphics[width=.45\textwidth,clip]{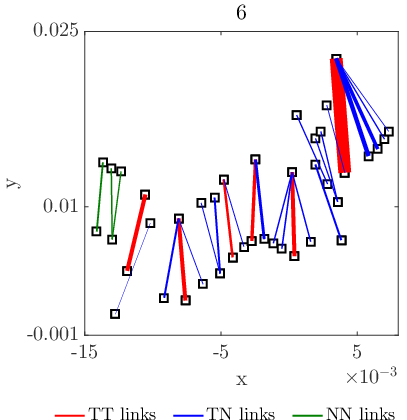}}
    \put(5,0){\includegraphics[width=1\textwidth,clip]{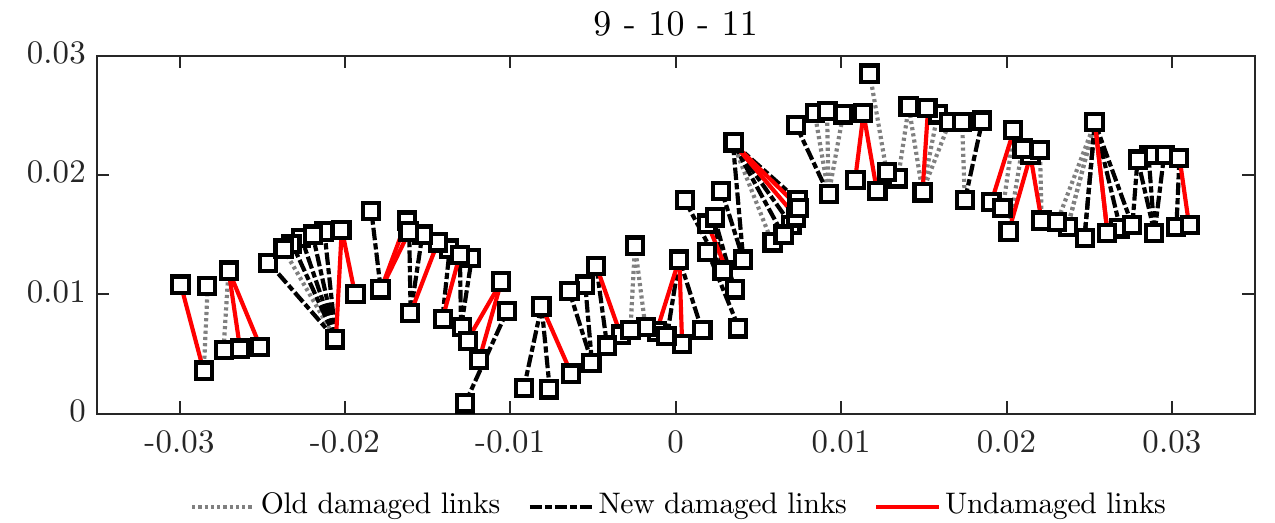}}
  \put(-5,550){\small{(a)}}
  \put(-5,375){\small{(b)}}
  \put(230,375){\small{(c)}}
  \put(-5,165){\small{(d)}}
  \put(148,425){\line(-2,-1){20}}
  \put(110,402){\line(-2,-1){62}}
  \put(272,425){\line(-3,-2){40}}
  \put(223,390){\line(-3,-2){30}}
\end{picture}
\caption{(Color online) {\it Heightened interdependency among contacts in $B^{*}$ predispose them to cascading failure}. (a) Grains colored red (blue) belong to the upper (lower) portion of the specimen in Data I.  Contact types for the highlighted region in (a) at: (b) stage 2 and (c) stage 6. Recall that Stage 2 is when the tensile force chain network is first established and stages 2-6 see a steady increase in the applied tensile load. Line thickness is proportional to the magnitude of the contact force. (d) Evolution of bonds as failure cascades in $B^{*}$ across stages 9-11 in Data I.  Similar trends apply to Data II (not shown).} 
\label{fig:mcSeg}      
\end{figure}

Despite a predisposition to force congestion, we have seen how bottlenecks interact to curtail damage in the dominant bottleneck $B^*$.  Here we demonstrate cooperative behavior among the contacts in $B^*$ with the same effect.  In both specimens, we observe forces to be spread out across member contacts in $B^*$, such that damage is confined to low capacity links in the initial stages (Figure~\ref{fig:evolTT}) --- leaving behind a web of mostly strong contacts to support the tensile force chains prior to peak load (Figure~\ref{fig:mcSeg} and Section 5 of \cite{DiB2018}).

In Data I, significant redistributions in the contact forces take place as early as stages 2-4, before the onset of damage in stage $5$: non-identity transitions frequently occur and the most common are between neighbors of force chains, NN $\rightarrow$ TN and TN $\rightarrow$ NN (Section 5 of \cite{DiB2018}) and Figure~\ref{fig:evolTT}). Transitions TT $\rightarrow$ TN and TT $\rightarrow$ NN are relatively rare prior to peak load, suggesting that tensile force chains in $B^*$ endure the pre-failure regime.

\begin{figure}[H]
\centering
\begin{picture}(420,430)(0,0)
  \put(5,215){\includegraphics[width=1\textwidth,clip]{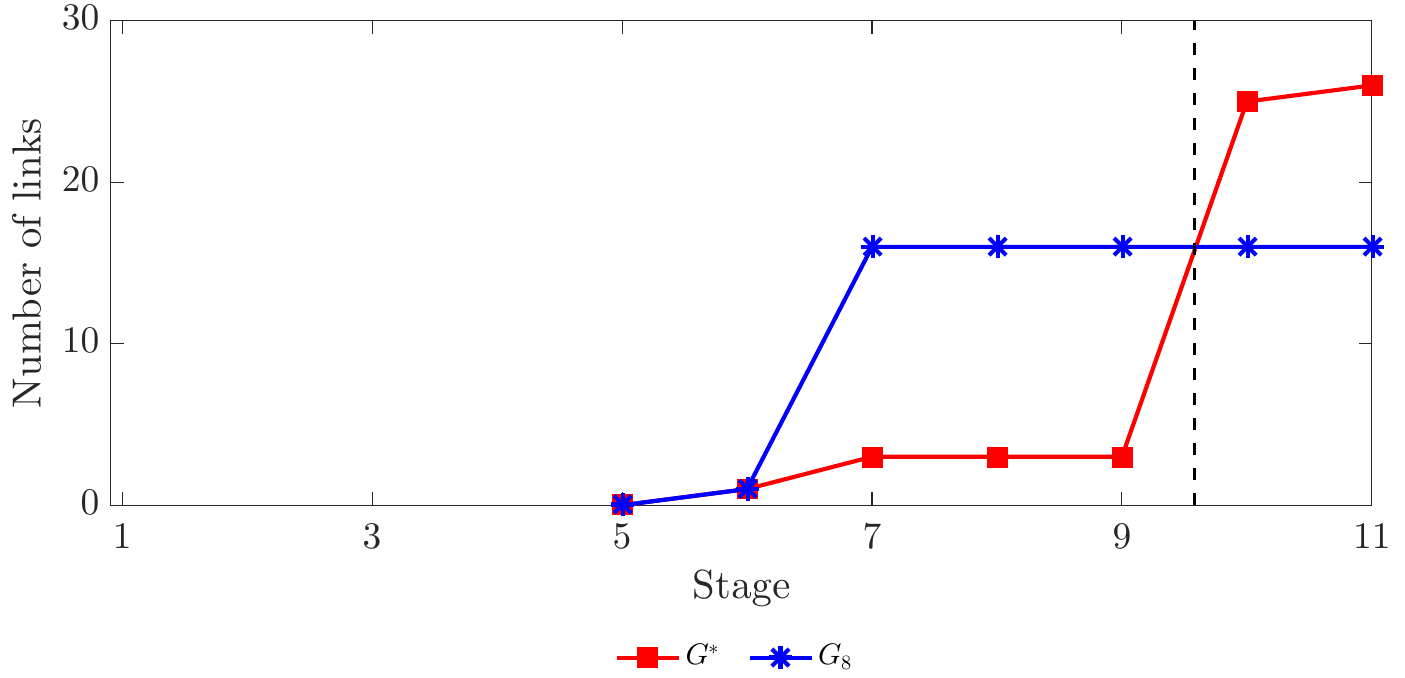}}
 \put(45,273){\includegraphics[width=.4\textwidth,clip]{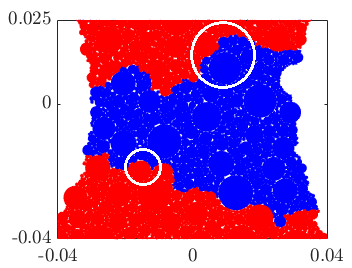}}
 \put(5,0){\includegraphics[width=1\textwidth,clip]{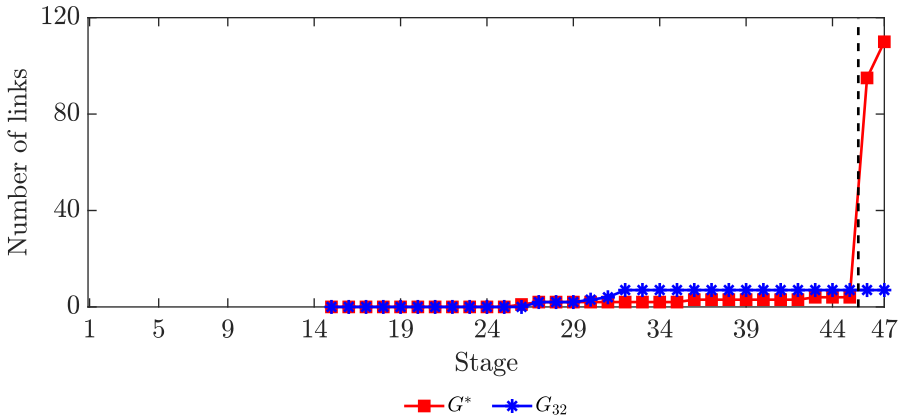}}
 \put(55,59){\includegraphics[width=.33\textwidth,clip]{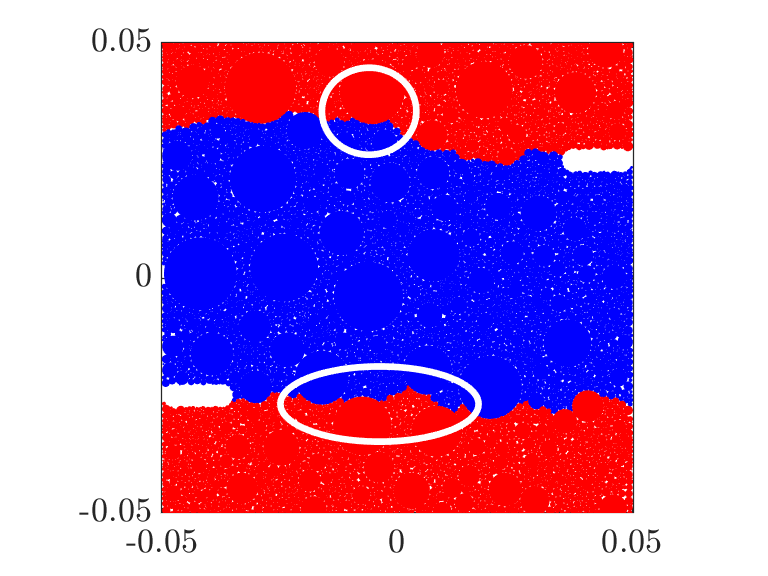}}
     \put(-5,420){\small{(a)}}
  \put(-5,205){\small{(b)}}
\end{picture}
\caption{(Color online) {\it Suppressed failure followed by cascading failure in the primary force bottleneck.}  Crack interaction as described by the evolution of the size (number of links) of the two largest components of the damage network $D_\mathcal{N}$ for (a) Data I and (b) Data II. Dashed vertical line marks stage at peak load. Inset (a): general location at stage 11 of $G_{8}$ (lower circle) and $G^{*}$ (upper circle) shown with the primary and secondary bottlenecks (red-blue interfaces).  Inset (b): general location at stage 47 of $G_{32}$ (upper circle) and $G^{*}$ (lower circle) shown with the primary and secondary bottlenecks (red-blue interfaces). }
\label{fig:cascade}     
\end{figure}

In Data II, the high level of redundancy in $\mathcal{N}$ leads to a more robust specimen.  Pre-failure transitions in $B^*$ are rare and occur only when bonds at TN contacts that support tensile force chains break (Section 5 of \cite{DiB2018}). 

Further evidence of a restrained damage in the primary bottleneck $B^{*}$ in the pre-failure regime can be found in the evolution of the damaged network $D_\mathcal{N}$ for both specimens (Figure~\ref{fig:cascade}).  In $D_\mathcal{N}$, the links represent the broken bonds and the nodes are the damaged grains, i.e., those with at least one broken bond ( Section 7 of \cite{DiB2018}).  In the lead up to peak load, the largest and second largest connected components of $D_\mathcal{N}$, $G_{8}$ (or $G_{32}$) and $G^{*}$, respectively lie in the region of the bottlenecks $B_{8}$ (or $B_{32}$) and $B^{*}$, consistent with the damage maps in Figure~\ref{fig:ngd}.  

In particular, in Data I, the earlier result that $B_{8}$ initially sustains most of the damage (Figure~\ref{fig:arb} (b)) is consistent with $G_{8}$ being larger than $G^{*}$ across stages 5-9.  However, as tensile forces increase towards the bond capacities, a critical point is reached when even a small increase in force triggers a cascade of bond failures across multiple interconnected contacts in $B^{*}$ (inset Figure~\ref{fig:cascade}).  Consequently, $G^{*}$ outgrows $G_{8}$ and becomes the giant component in the ensuing stages 10-11 in the failure regime. This explains why the residual strength of this specimen is mainly due to the $\mathcal{P}$ links on the left side of the specimen: the relatively sparse $\mathcal{P}$ links on the right side coincides with $G^{*}$ being on the right side of $B^{*}$ (stage 11, Figure~\ref{fig:ptfcs} (a)).  All of these trends similarly hold for Data II, except that $G^{*}$ is on the left side of $B^{*}$, consistent with the highly degraded and sparse $\mathcal{P}$ pathways on the left side of the specimen during failure (stage 47, Figure~\ref{fig:ptfcs} (b)).


\section{Discussion}
\label{sec:sum}

Damage impairs pathways for force transfer.  In turn, this disrupts force transmission and weakens a material's robustness against future damage under load.  With {\it a priori} knowledge of the bond strengths and connectivities in the contact network, we uncovered an optimized force transmission, characterized by two defining transmission patterns: the optimized flow routes (Figure \ref{fig:ptfcs}) and the force bottleneck (Figure \ref{fig:minCut}).  Both have a capability to predict and explain the pre-failure propagation of force and damage, from the microscopic to the macroscopic scale.  We now discuss the implications of key findings and how these can help resolve some open questions on the fracture mechanics of quasi-brittle granular materials.

A hallmark of force transmission in granular media are force chains.  The recent surge of interest in these emergent structures from studies of quasi-brittle and cohesive materials (e.g., rock~\cite{burnley2013}, gels \cite{SCHENKER20081443}, wheat endosperm \cite{TOPIN2009215}, sediments \cite{jiang2014bond}, ice \cite{POLOJARVI201512}, concrete~\cite{nitka2015}, magmas \cite{bergantz2017kinematics}, cement paste \cite{SUN201769}, asphalt \cite{Chang_2017}) --- concerns not just the pattern they form but also the nature of the self-organization process behind their formation. Unravelling this process may yield clues to the question raised in~\cite{burnley2013}: how and why do these preferential paths for force transmission arise?  Here we found that the optimized flow routes $\mathcal{P}$ both thread through and predict the majority of tensile force chains, thus giving insight into the self-organization rule that governs their formation.  Force chains in $\mathcal{P}$ are essentially the ``highways'' in an optimized force transport network $\mathcal{N}$ (Figure~\ref{fig:ptfcs}).  They emerge in those routes that can transmit the global force transmission capacity $F^*$ through shortest possible, percolating paths in the direction of the major (here, most tensile) principal stress.  In systems with high redundancy, these critical load-bearing pathways are protected on both sides by alternative force pathways that can share and take the load off, as well as replace damaged, force chains.

With respect to failure evolution, studies have generally looked to sites of high stress concentration when searching for the origin of macrocracks (e.g., \cite{Gu2013}).  This makes force chains obvious suspect locales for incipient failure.  However, force chains also have comparatively high capacities (Figure \ref{fig:evolTT}).  Thus findings here suggest a more nuanced ``systems'' approach is needed, which recognizes that proximity to failure matters in an optimized transmission process.  That is, the sites that are the most susceptible to force congestion and damage are those which not only transmit high stresses but have the smallest capacities.

It is clear from our results that force bottlenecks control the onset and propagation of macrocracks (Figure \ref{fig:minCut}).  The bottleneck is an emergent property of the whole network $\mathcal{N}$ -- not possessed by individual contacts nor by individual force chains.  In particular, the bottleneck is a vulnerable and critical ``nerve-centre'' of $\mathcal{N}$: (a) its member contacts, all of which lie in $\mathcal{P}$, mostly transmit above-average forces while having the least total capacity; (b) it is the site where macrocrack ultimately forms; and (c) its capacity controls the global transmission capacity (Figure~\ref{fig:arb}). We observed two bottlenecks in the pre-failure regime.  The first is the bottleneck at the initial undamaged state of the specimen.  Persisting in the same location of the specimen across multiple states of the pre-failure regime, this primary bottleneck gives an accurate and early prediction of the primary (dominant) macrocrack that develops in the failure regime.  The second bottleneck emerges at a state close to peak load and is where the secondary macrocrack emerges.  Past studies have raised the question on whether or not the initiation point of a crack that leads to failure can be predicted from known microstructural features (e.g., location of the flaw~\cite{Gu2013}).   That the bottlenecks distinguish themselves by having capacities far lower than those of other cuts or partitions of the specimen -- even before the onset of damage -- suggests this may be possible (Figure~\ref{fig:arb}).


A question that now emerges is: what role do force chains play in the evolution of failure? To answer this, we turned to recent studies that have underscored the importance of stress redistributions on the evolution of progressive fracture (e.g., \cite{LUO2017264,berthier2017damage}).  Our findings concur with this view.  In the nascent stages of the pre-failure regime, stress redistributions do play a critical role (Figure~\ref{fig:rerouted}).  We discovered a two-pronged cooperative mechanism that underlies robustness.  This mechanism, enabled by the available pathway redundancies, maximizes global transmission capacity.  In the first prong, bottlenecks interact cooperatively: bottlenecks take turns in accommodating damage to minimize the unavoidable reduction in global transmission strength.  In the second prong, contacts in the primary bottleneck similarly interact cooperatively by spreading and sharing the forces to induce the same effect.  Attendant pre-failure damage in the primary bottleneck is confined essentially to below average capacity member bonds whose breakage: (a) incurs a comparatively low reduction in the bottleneck capacity (and, in turn, the global transmission capacity); and (b) leaves behind a web of mostly strong contacts to support the tensile force chains in the region, thus curtailing their failure despite a predisposition to force congestion.  But all these come at a cost: a heightened interdependency among the dominant bottleneck contacts in the final stages of the pre-failure regime.  Just before peak load, a critical point is reached when even a small increase in force triggers a cascade of bond breakages in this bottleneck, in turn precipitating catastrophic global failure (Figures~\ref{fig:evolTT}-\ref{fig:cascade}).  Thus, the mechanism uncovered here --- though initially mitigates damage in the dominant bottleneck --- elicits the opposite effect.

Finally, we raise the limitations of this analysis and highlight where future research may be directed.  The optimized force transmission process uncovered in this study can be subjected to many other factors.  Among these are loading conditions that give rise to interactions among tensile and compressive force chains~\cite{Cho2007}, as well as network rewiring where new contact paths for force transmission emerge from grain rearrangements (e.g., in specimens under confined compression~\cite{SUN201769,jiang2014bond}).  These warrant further research as they could lead to patterns of evolution different from those reported here.  Note also that the present formulation accounts only for the one microscale failure mechanism that was observed in the pre-failure regime (i.e., bond breakage in tension).  There are various strategies for extending this analysis to accommodate the influence of other forms of heterogeneities and failure mechanisms on force and fracture propagation.  For example, heterogeneities at the sub-grain level can be addressed by modeling each grain as a sub-network of nodes and links, similar to past work on grain fracture \cite{tordesillas2015network}, with sub-grain links given different capacities.  Heterogeneities at the grain-grain contact level have been addressed here with both material and geometrical (grain sizes) properties influencing the tensile strength of each bonded contact in the 3-phase specimen of Data II.  To account for additional sources of strengths, each contact may be modeled by multiple links. For example, a bonded contact may be given two links: one whose capacity reflects the shear strength of the bond, while the other the tensile strength.  Last but not least, while all these strategies concern force transmission, in principle, the framework developed here can be applied to study the conductivity of other mechanical properties in heterogeneous media, such as interstitial pore fluid \cite{SciRep}, heat and energy, given data on the available pathways for transmission and their relative capacities.

\section{Conclusion}
\label{sec:conclude}

A framework that leverages microstructural data assets, be they from experiments or physics-based models, has been developed to examine the interdependent evolution of damage, force transmission and robustness in heterogeneous, quasi-brittle granular media.  Using this multiscale framework, we demonstrated that data on the internal strengths and connectivity of a system can be mapped to an evolving complex flow network, from which nontrivial patterns in the dynamics on and of this network can be extracted to gain important fundamental insights on transmission processes in the presence of disruptions.  This study opens the door for other applications of network flow, specifically, in fundamental studies of multiscale processes involving the transmission of interstitial pore fluid, heat, energy, force, kinematics, stress, strain, etc. in granular materials as well as other forms of heterogeneous media.  In practical settings, our approach may prove useful in {\it de novo} design of mechanically robust aggregate and composite materials through rational fine-tuning of the heterogeneities in microstructural fabric and strength. Finally, this study also casts new light on the dynamics of critical bottlenecks as precursors for endogenous cascading failures and, as such, may have implications for other complex transmission systems such as infrastructure and communication networks.

\section*{Acknowledgement}
This work was supported by grants to AT from the US Army Research Office (W911NF-11-1-0175) and the US Air Force (AFOSR 15IOA059). MN and JT were supported under the project: ‘‘‘Experimental and numerical analysis of coupled deterministic-statistical size effect in brittle materials” financed by the National Science Centre NCN (UMO-2013/09/B/ST8/03598).

\section*{References}
\bibliography{References}

\newpage
\appendix

\section{Symbols and nomenclature}
\label{sec:symbols}
The list below contains the symbols and nomenclature used in Section 4.  It is divided into two groups, each arranged in alphabetical order: symbols in the Greek alphabet, and symbols in the English alphabet.
\begin{flushleft}
\begin{longtable}{p{.20\textwidth}  p{.80\textwidth}} 
\hline
\hline
\hline
$\alpha$ & Number of replacement links to which flow is diverted\\
$\gamma$ & Number of links that cease to be part of $\mathcal{P}$\\
$\d^{-}(v)$ & Arcs entering node $v$\\
$\d^{+}(v)$ & Arcs leaving node $v$\\
$\d^{-}(s)$ & Arcs entering the source or supersource $s$\\
$\d^{+}(s)$ & Arcs leaving the source or supersource $s$\\
$\d^{+}(S)$ & Cut of $G$ induced by $S$\\
$\rho$ & Ratio of the number of links in $\mathcal{P}$ relative to its value prior to damage\\
\hline
\hline
$B$ & Minimum cut\\
$B_{min}$ & Minimum edge cut\\
$b$ & Demand function\\
$b_v$ & Demand of $v$\\
$c_e$ & Cost of $e$\\
$c$ & Cost function\\
$E$ & Set of arcs (contacts) of $G$\\
$e$ & Arc or directed link \\
$\mathcal{F}$ & Flow network\\
$\mathcal{F}_1$ & Flow network with unit link capacities\\
${F}_1$ & Maximum flow on $\mathcal{F}_1$\\
\textbf{$\text{F}_{\text{ji}}$} & Normal tensile force acting on grain $\textbf{i}$ imposed by grain $\textbf{j}$\\
$F^*$ & Maximum flow\\
$f(x)$ & Net flow transmitted from $s$\\
$G$ & Directed network of $\mathcal{N}$\\
$\mathcal{N}$ & Bonded contact network\\
$\mathcal{P}$ & Optimized flow routes\\
$p_{min}$ & Pathway redundancy\\
$\mathbb{R}$ & Real numbers\\
$\mathbb{R}_{+}$ & Non-negative real numbers\\
$R_{re}$ & Reroute score\\
$S, T$ & Disjoint set of nodes (grains) attached, respectively, to the supersource $s$ (top wall) and supersink $t$ (bottom wall)\\
$s$ & Source, supersource\\
$t$ & Sink, supersink\\
$u$ & Capacity function for all of $E$\\
$u_e$ & Capacity of $e$\\
$u(\d^{+}(S))$ & Capacity of $\d^{+}(S)$\\
$V$ & Set of nodes (grains) of $G$\\
$v, w$ & Nodes representing grains $v$, $w$\\
$(v,w)$ & Link between nodes $v$ and $w$\\
$x$ & Feasible $(s,t)$-flow, feasible $(S,T)$-flow\\
$x_e$ & Flow on $e$\\
\hline
\hline
\hline
\end{longtable}
\end{flushleft}



\end{document}